\documentclass[twocolumn,aps,prx,floatfix,reprint,showpacs,preprintnumbers,superscriptaddress,amsmath,amssymb,nofootinbib]{revtex4-1}
\usepackage{color} 
\usepackage{units}
\usepackage{mathtools}
\usepackage{bm}
\usepackage{amsmath}
\usepackage{amssymb}
\usepackage{stmaryrd}
\usepackage{graphicx}
\usepackage[unicode=true,pdfusetitle,
 bookmarks=true,bookmarksnumbered=false,bookmarksopen=false,
 breaklinks=false,pdfborder={0 0 1},backref=false,colorlinks=true]
 {hyperref}\hypersetup{citecolor=blue,linkcolor=blue,urlcolor=blue}
\usepackage[capitalize]{cleveref}

\usepackage{dcolumn}
\usepackage{bm}
\usepackage{multirow}
\usepackage[dvipsnames]{xcolor}
\usepackage{enumerate}
\usepackage{float}
\usepackage{units}
\usepackage{soul}

\newcommand{\TRuu}{\overrightarrow{T}_{\uparrow\uparrow}}
\newcommand{\TRdd}{\overrightarrow{T}_{\downarrow\downarrow}}
\newcommand{\TRud}{\overrightarrow{T}_{\uparrow\downarrow}}
\newcommand{\TRdu}{\overrightarrow{T}_{\downarrow\uparrow}}

\newcommand{\TR}{\overrightarrow{T}}
\newcommand{\TL}{\overleftarrow{T}}
\newcommand{\RR}{\overrightarrow{R}}
\newcommand{\RL}{\overleftarrow{R}}
\newcommand{\TRu}{\overrightarrow{T}_{\uparrow}}
\newcommand{\TRd}{\overrightarrow{T}_{\downarrow}}

\newcommand{\RRu}{\overrightarrow{R}_{\uparrow}}
\newcommand{\RRd}{\overrightarrow{R}_{\downarrow}}
\newcommand{\RLu}{\overleftarrow{R}_{\uparrow}}
\newcommand{\RLd}{\overleftarrow{R}_{\downarrow}}

\newcommand{\TRmat}{\overrightarrow{\underline{T}}}
\newcommand{\TLmat}{\overleftarrow{\underline{T}}}
\newcommand{\RRmat}{\overrightarrow{\underline{R}}}
\newcommand{\RLmat}{\overleftarrow{\underline{R}}}
\newcommand{\Rmat}{\underline{\cal R}}
\newcommand{\Imat}{\underline{I}}

\newcommand{\GR}{\overrightarrow{G}}
\newcommand{\GL}{\overleftarrow{G}}

\newcommand{\jRu}{\overrightarrow{\jmath}_{\uparrow}}
\newcommand{\jRd}{\overrightarrow{\jmath}_{\downarrow}}
\newcommand{\jLu}{\overleftarrow{\jmath}_{\uparrow}}
\newcommand{\jLd}{\overleftarrow{\jmath}_{\downarrow}}

\newcommand{\PL}{\overleftarrow{\Phi}}
\newcommand{\PR}{\overrightarrow{\Phi}}

\newcommand{\dz}{\delta{z}}
\newcommand{\dt}{\delta{t}}

\newcommand{\emax}{\epsilon_{\rm max}}

\newcommand{\dd}{{\rm d}}

\begin{document}
\preprint{APS/123-QED}
\title{Theory of superdiffusive spin transport in noncollinear magnetic multilayers}
\author{Pavel Bal\'a\v z}
\email{balaz@karlov.mff.cuni.cz}
\affiliation{FZU -- Institute of Physics of the Czech Academy of Sciences, Na Slovance 1999/2, 182 21 Prague 8, Czech Republic}
\author{Maciej Zwierzycki}
\affiliation{Institute of Molecular Physics, Polish Academy of Sciences, Smoluchowskiego
17, 60-179 Pozna\'{n}, Poland}
\author{Francesco Cosco}
\affiliation{Department of Physics and Astronomy, Uppsala University, Box 516,
SE-75120 Uppsala, Sweden}
\author{Karel Carva}
\affiliation{Charles University, Faculty of Mathematics and Physics, Department
of Condensed Matter Physics, Ke Karlovu 5, CZ 121 16 Prague, Czech
Republic}
\author{Pablo Maldonado}
\affiliation{Department of Physics and Astronomy, Uppsala University, Box 516,
SE-75120 Uppsala, Sweden}
\author{Peter M.\ Oppeneer}
\affiliation{Department of Physics and Astronomy, Uppsala University, Box 516,
SE-75120 Uppsala, Sweden}
\date{\today}
\begin{abstract}
Ultrafast demagnetization induced by femtosecond laser pulses in thin metallic layers 
is caused by the outflow of spin-polarized hot electron currents describable by the superdiffusive transport model.
These laser-generated spin currents can cross the interface into another magnetic
layer and give rise to magnetization dynamics in magnetic spin valves with noncollinear magnetizations. 
To describe ultrafast transport and spin dynamics in such nanostructures we develop here the superdiffusive theory for general noncollinear magnetic multilayers. Specifically, we introduce an Al/Ni/Ru/Fe/Ru multilayer system with  noncollinear Ni and Fe magnetic moments and analyze how the ultrafast demagnetization and spin-transfer torque depend on the noncollinearity. 
We employ \emph{ab initio} calculations to compute the spin- and energy-dependent transmissions of hot electrons at the interfaces of the multilayer. 
Taking into account multiple electron scattering at interfaces and spin mixing in the spacer layer   
we find that the laser-induced demagnetization of the Ni layer and magnetization change of the Fe layer strongly depend on the angle between their magnetizations. Similarly, the spin-transfer torques on the Ni and Fe layers and the total spin momentum absorbed in the Ni and Fe layer are found to vary markedly with the amount of noncollinearity. 
 These results suggest that changing the amount of noncollinearity in magnetic multilayers
one can efficiently control the hot electron spin transport, which may open a way toward achieving fast, laser-driven spintronic devices.
\end{abstract}
\pacs{}

\maketitle

\section{Introduction}

\label{Sec:Intro}

Fast and energy-efficient control of the 
magnetic configuration of nanostructures is a fundamental requirement for practical spintronic devices.  
Ever since the discovery of the spin-transfer torque (STT) by
Slonczewski~\citep{Slonczewski1996} and Berger~\citep{Berger1996} in
1996, the STT has been seen as a major candidate for achieving this goal.
The main obstacle for widespread application of STT are the high
current densities required for effectuating the change of magnetization 
alignment. One possible way of addressing this problem is to use the so called spin-orbit torque (SOT)~\citep{Miron:Nature2011,Liu:Science2012,Jhuria2020_SOTsw_ps}, 
a torque variant where the
pure spin current penetrating into the magnetic layer is generated by the
spin Hall effect (SHE)~\citep{Hirsch:PRL1999} or related effects as the Rashba-Edelstein effect originating microscopically from
strong spin-orbit coupling~\citep{Sinova:RevModPhys2015}.

Another possibility to manipulate spins has emerged from the line of research which started with
the breakthrough experiment by Beaurepaire \emph{et al.}~\citep{Beaurepaire1996}, who observed ultrafast
demagnetization of a nickel thin film after exposing it to a
femtosecond laser pulse. The intriguing possibility of sub-picosecond
scale manipulation of the magnetic 
order inspired a number of
follow-up studies, both experimental and theoretical \cite{Carva2017_Handbook}. However, it took
more than 10 years until research in the field of ultrafast
laser-induced magnetization processes had been combined with
spintronics leading to discoveries of optical spin transfer
torque~\citep{r_12_Nemec_opticalSTT} and optical spin-orbit
torque~\citep{r_13_Tesarova_Nemec_Observ_SOT}.  The importance of spin-transfer effects in the demagnetization of metallic multilayers has been observed by Malinowski~\emph{et  al.}~\citep{Malinowski2008:NatPhys}. 
The demagnetization in metallic structures was attributed to the  spin transfer of laser-excited hot electrons 
moving across the sample. Since electron velocities and life times in magnetic metals are strongly spin-dependent, electronic motion leads
to fast spin currents contributing to ultrafast
demagnetization. Theoretically, this process has been described by the superdiffusive spin-dependent transport model of
Battiato, Carva, and Oppeneer~\citep{Battiato2010,Battiato2012}, explained in more 
detail in Sec.~\ref{Sec:Superdiff}. The important role of spin currents in ultrafast demagnetization is supported by a number of experimental
observations~\citep{Melnikov2011,Turgut2013,Schellekens2014,Choi2014,Bergeard2016,Hofherr2017,Malinowski2018,Kumberg2020,Kuhne2022,Iihama2018_SingleShot_AOS_HotEl,Remy2020_UltraSC_SingleSwitch}.

Additional support for the nonthermal hot electron spin-current picture comes from
Ref.~\citep{Rudolf2012}, where the effect of a femtosecond laser pulse
has been studied in magnetic trilayers, consisting of two
ferromagnetic layers separated by a nonmagnetic spacer. When the
magnetizations of the ferromagnetic layers are collinearly aligned and one of them is
illuminated by the laser pulse, its ultrafast demagnetization is
followed by the subsequent transient increase in the magnetization of
the second layer. However, in the antiparallel configuration the same procedure
leads to a decrease in the second layer's magnetization.  This effect can
be readily understood in terms of the hot electron spin current being
generated in the first layer and carrying the spin polarization aligned
with this layer into the second one. Moreover, the results were
shown to be in a good agreement with the superdiffusive transport
model. On the other hand, Eschenlohr {\em  et~al.}~\citep{Eschenlohr2017}
have inspected the same kind of trilayer using L-edge x-ray magnetic spectroscopy and found no enhancement of
magnetization. 

The discrepancy between experimental results calls 
for further development of the theoretical
approach allowing for even more precise modelling
of hot electron transport in complex multilayered structures.
In the current manuscript we discuss the following twofold improvement to the superdiffusive transport model: (i) the inclusion of a realistic, material-specific description of the scattering at the interfaces and 
(ii) the extension of the formalism to the case of noncollinear magnetic configurations.

Noncollinear magnetic alignments appear frequently, adding a novel aspect to ultrafast spin transport in magnetic nanostructures. 
For example, a number of recent experiments studied ultrafast demagnetization in multilayers combining a ferromagnet and Pt
\cite{Siegrist2019_Nat_LightMag_CohControl, Vaskivskyi2021_CoPt,Hennes2022_ElemSelDemag_CoPt}. In the case of Co/Pt it is possible 
to achieve not only  in-plane, but also out-of-plane magnetization orientation \cite{Hennes2022_ElemSelDemag_CoPt}, which makes such
system an interesting candidate for studies of the role of noncollinearity. Several recent experiments probed the ensuing 
magnetization dynamics in noncollinear magnetic multilayers \citep{Schellekens2014,Choi2014,Carva2014_NPhys,Razdolski2017,Lalieu2019_THz_SpW_Noncol}.

The most significant manifestation of the optically induced spin currents in
noncollinear magnetic multilayers or magnetic textures is the appearance of a
STT leading to magnetization dynamics.
To this end, a number of experiments focused on magnetic trilayers with noncollinear
(usually perpendicular) magnetizations. These experimental observations ~\citep{Schellekens2014,Choi2014,Razdolski2017,Lalieu2019_THz_SpW_Noncol} and theoretical
models~\citep{Balaz2018:JPCM} 
suggest laser excitation of hot electrons in one magnetic layer can
lead in a quick succession to small angle precession of the moment of the second magnetic layer, caused by the STT 
due to the spin current, even though both magnetic
layers are separated and magnetically decoupled by a nonmagnetic spacer.
Interestingly, experiments~\citep{Lalieu2017,Razdolski2017} and
numerical simulations~\citep{Ulrichs2018,Ritzmann2020:PRB} have
revealed that a spin current of hot electrons in noncollinear spin
valves can trigger terahertz spin waves. 

The interplay between the
laser-induced spin currents and localized magnetization 
touches upon a central interest in ultrafast spintronics,  having the potential to fulfil both requirements mentioned earlier in  
this section -- speed and efficiency.
The first characteristic results from the fundamentally different transport regime, distinct from
the usual diffusive/ohmic one, governing the propagation of the spin
information across the system. The hot electrons propagate initially in nearly
ballistic fashion, which means that 
their real velocity (Bloch velocity) is relevant, which is many orders of magnitude higher than the
drift velocity characterizing the collective motion in diffusive
systems.  The efficiency follows from the cascade effect~\citep{Battiato2010,Battiato2012}
where after the absorption of a
single photon, the excited electron can, in subsequent scatterings,
excite many more, multiplying thereby the number of particles 
carrying the spin information~\citep{Battiato2012}.

To describe laser-induced STT and magnetization dynamics in magnetic multilayers, 
we developed recently an effective model for the STT induced by hot electrons
in perpendicular noncollinear spin valves~\citep{Balaz2018:JPCM}. Moreover, by means
of atomistic spin dynamics we have demonstrated the possibility of
generation of THz spin waves in a magnetic film due to the superdiffusive STT
exerted by hot electrons~\citep{Ritzmann2020:PRB}. Later on, to
describe arbitrary noncollinear magnetic textures like a domain wall, we
generalized the superdiffusive transport model by including a spin rotation
transformation~\citep{Balaz2020:PRB}
and predicted a high speed shift of the domain wall center due to a femtosecond laser pulse.
The model was also used to predict optimal spacer thickness to maximize the hot electron spin current 
and an impressive agreement between these theoretical predictions and experiment 
has been observed~\cite{Bergeard2020_Tailor_HotEl}. 

As a further motivation,  there is an ongoing effort to achieve magnetization reorientation into a desired direction by means of a light pulse. One example is the precession caused by the Zeeman interaction with the magnetic component of light, an effect best manifested in the THz range \cite{Shalaby2018_co_CohIncohDyn_THz}. Another case is the laser-induced magnetization due to the inverse Faraday effect \cite{Berritta2016}. In both these cases an optical pulse will cause inevitable excitation of electrons in addition to the magnetization reorientation, and superdiffusive currents may thus occur. The resulting demagnetization and STT could help ease the switching, on the other hand the loss of magnetization could unfortunately dominate over switching. It is therefore needed to provide the description of superdiffusion related effects also for arbitrary angles reached during the switching process in order to describe magnetization dynamics of such process completely.  
Interesting discoveries regarding magnetization dynamics in noncollinear structures with a chirality were made recently \cite{Kerber2020_ChiralMagOrder_OptExc}.
 Noncollinear antiferromagnetic spin structures not restricted to perpendicular angles were also predicted to feature unusual properties of the current-induced STT, including a self-generated torque \cite{Ghosh2022_STT_Ncol_AFM}.  

In the current paper, we study the hot electron transport in the
magnetic multilayer, namely the Al/Ni/Ru/Fe/Ru structure also studied
experimentally \citep{Rudolf2012,Turgut2013,Eschenlohr2017}. Using the
superdiffusive transport model, we inspect the demagnetization as well
as spin-transfer torques acting on the magnetic layers after the hot
electrons are excited by a femtosecond laser pulse. In order to
properly describe the transport through the interfaces we employed the
{\em ab initio} wave function matching(WFM) method
\citep{Xia:prb06,Zwierzycki2008:PSSb} to calculate 
spin- and energy-dependent transmissions of hot electrons.  Thus, the
superdiffusive transport model allows us to account for multiple
reflections of electrons from the interfaces. In case of noncollinear
magnetic configuration, spin mixing in the central nonmagnetic layer
has to be taken into account. Since this layer is usually just a few
atomic monolayers thick, we model the nonmagnetic spacer as a single
interface between the magnetic layers allowing spin mixing of the hot
electrons passing through it.  In collinear magnetic configurations~\cite{Schellekens2014},
we have shown that our assumption has just a minor effect on the
results obtained by the superdiffusive transport model.

The importance of interface scattering is well established for 
the case of conventional Fermi-level spin transport \cite{Bass:jmmm99,Bass:jpcm07}. Its effect on the
process of ultrafast demagnetization in a ferromagnet/normal 
metal bilayer was recently studied independently, and found to be meaningful,
in Ref.~\citealp{Lu:PRB2020} using an approach similar to that
outlined in Secs.~\ref{Sec:Superdiff} and \ref{Sec:Int_method}.

The paper is organized as follows. In Section~\ref{Sec:Superdiff} we
present the main features of the spin-dependent superdiffusive transport model. 
The system for which the calculations are performed is described in Sec.~\ref{Sec:Model} and the method 
for calculating the scattering coefficients for interfaces is discussed in Sec.~\ref{Sec:Int_method}.
In Sec.~\ref{Sec:Int_results} we present our calculations of transmissions through the interlayer interfaces.
The results for ultrafast demagnetization
and spin-transfer torque in a noncollinear Ni/Ru/Fe trilayer are described
in Sec.~\ref{Sec:Demag} and Sec.~\ref{Sec:STT} respectively. We conclude in Sec.~\ref{Sec:Conclusions}.

\section{Theory}

\subsection{The superdiffusive transport model}
\label{Sec:Superdiff}

The superdiffusive spin-dependent transport model~\citep{Battiato2010,Battiato2012} has 
been developed to describe the local loss of magnetic moment of metallic samples 
after they are exposed to a laser pulse.
In our study, a laser pulse initiates the electronic transport by exciting 
electrons with energies at and below the Fermi level -- typically occupying strongly
$d$-hybridized and relatively immobile states -- into higher energy bands of $sp$ 
character. These $sp$-electrons are characterized by higher velocities, hence, they behave as 
itinerant particles moving through the sample~\citep{Battiato2014_JAP}. We note that an alternative approach to describe the collision-rich motion of excited hot electrons in metallic layers has been developed on the basis of Boltzmann transport theory \cite{Nenno2016,Nenno2018}.

The superdiffusive transport model describes two spin channels, 
tagged as $\sigma\in\{\uparrow,\downarrow\}$,  for the electronic transport. 
Each spin channel is characterised by the hot electron velocities,  $v_{\sigma}(\epsilon)$, and lifetimes,  $\tau_{\sigma}(\epsilon)$,  which depend on the electron energy $\epsilon$,
and,  in magnetic materials,  on the spin $\sigma$ as well.  
Because of the distinct transport properties of the two spin channels, the current of hot electrons gets 
spin polarized in the magnetic layer.
In addition, in the case of multilayers,  spin filtering
via multiple spin-dependent transmissions and reflections at the interfaces further 
contributes to the spin polarization of the current~\citep{Battiato2014_JAP}.

The outflow of the spin-polarized current 
initiated by a laser pulse results in a loss of local magnetic
momentum and demagnetization.  Due to the high velocities of hot electrons,  
the demagnetization happens on a timescale of hundreds of femtoseconds~\cite{Battiato2010}. 
Initially, the hot electrons' propagation is ballistic; however, due to the relaxation processes caused by electron scattering,  it continuously changes its character.  
In about $500\,{\rm fs}$ up to $1\,{\rm ps}$ it develops into a purely diffusive one.  
In the time interval up to 500 fs,  
the transport proceeds in the superdiffusive regime~\citep{Battiato2012}.

Let us describe briefly the main features of the superdiffusive spin-dependent
transport model~formulated in Refs.\,\citealp{Battiato2010,Battiato2012},
which revolves around the the equation  of motion for the  hot electron density $n_\sigma (\epsilon,z,t)$, with spin $\sigma$, energy $\epsilon$, and position $z$
\begin{equation}
\frac{\partial n_\sigma (\epsilon,z,t)} {\partial t}+ \frac {n_\sigma (\epsilon,z,t)}{\tau_\sigma (\epsilon,z)}=
 \left (  -\frac {\partial  } {\partial z}  \hat \phi+ \hat I\right) S_\sigma^{\mathrm {eff}} (\epsilon,z,t),
\label{eom}
\end {equation}
where $\hat \phi$ is the flux operator which contains the dependence from the electrons' velocities 
and describes interlayer  transmissions and reflections.  Furthermore,  $S_\sigma^{\mathrm {eff}} (\epsilon,z,t)$
is an effective source term  describing the laser induced excitation of spin-polarized hot electrons
and scattering events within the material (see the discussion below).

For practical calculations we adopt the discretized version of the formalism, described in Ref.~\citealp{Battiato2014_JAP}, 
with the space divided into computational cells of width $\delta z$. Time and energy are also sampled in finite steps 
of $\delta t$ and $\delta \epsilon$, respectively. 

When treating the multilayers, one should include the possibility of scattering at the interlayer interfaces. 
It is convenient to formally treat each computational cell as an individual layer and its boundaries as 
interfaces characterized by specific transmission probabilities.  Therefore,  until the end of the current 
section, the term interface will refer to interfaces between the computational cells, even though only some of 
these correspond to actual interfaces between different metallic layers with nontrivial scattering coefficients.

Following Ref.~\citealp{Battiato2014_JAP},  the solution  of Eq.\  \eqref{eom} on the discrete grid,  
averaged over the cell centered at $z_i$ and omitting, for sake of brevity, the energy index $\epsilon$, 
can be written as
\begin{widetext}
\begin{equation}
\begin{split}
n_{\sigma}(z_i,t+\dt)=\; & e^{-\dt/\tau_{\sigma}(z_i)}n_{\sigma}(z_i,t)+ S_\sigma^{\mathrm {eff}}(z_i, t+\dt)\;+\\
  &  \   \sum_{ \sigma'} \left [ \TR_{\sigma,\sigma'}(z_i^-) \,\PR_{\sigma'}(z_i^-,t)+
 \TL_{\sigma,\sigma'}(z_i^-)\,\PL_{\sigma'}(z_i^-,t)-
  \TR_{\sigma,\sigma'}(z_i^+)\,\PR_{\sigma'}(z_i^+,t)-\TL_{\sigma,\sigma'}(z_i^+)\,\PL_{\sigma'}(z_i^+,t) \right ],
\end{split}
\label{Eq:n-sol}
\end{equation}
\end{widetext}
where $z_i^{\pm}=z_i\pm\dz/2$ are the positions of the left and right interface, respectively,  for each
cell centered at $z_i$. $\overrightarrow{\Phi}_{\sigma}(z_i^{\pm},t)$ 
and $\overleftarrow{\Phi}_{\sigma}(z_i^{\pm},t)$, are the fluxes of right and left
moving particles with spin $\sigma$ going  through the interface $z_i^{\pm}$ of the cell at time $t$. 
Each interface is characterized by energy-dependent transmissions of electrons moving to the right,
$\overrightarrow{T}_{\sigma\sigma'}$, and to the left $\overleftarrow{T}_{\sigma\sigma'}$
with $\sigma$ and $\sigma'$ indicating spin after and before the
transmission, respectively.
Thus,  the 1{st} term in the sum represents the incoming fluxes
through the left interface of the cell, while the 2{nd} term
 represents the outgoing fluxes through the left interface.  In a similar fashion, 
the 3{rd} term stands for the outgoing fluxes through the right interface,  and the last term 
expresses the incoming fluxes through the right interface.
The first two terms of Eq.~\eqref{Eq:n-sol} describe the variation
of $n_\sigma$ due to the spin relaxation and effective source,  respectively. The latter is made of two contributions, i.e.\
\begin{equation}
S_\sigma^{\mathrm {eff}}(z,t+\delta{t})=S_{\sigma}^{\mathrm {ext}}(z,t+\delta{t})+S_{\sigma}^{{\rm p}}(z,t+\delta{t})\,,\label{Eq:gen_source}
\end{equation}
where $S_\sigma^{\mathrm {ext}}(z,t+\delta{t})$ describes the excitation of hot electrons by the laser pulse, and $S_{\sigma}^{{\rm p}}(z,t+\delta{t})$ is the term describing the effects of scattering, mostly caused by electron-electron interactions, calculated as
\begin{equation}
\begin{split}S_{\sigma}^{{\rm p}}(\epsilon,z,t+\delta{t})= & \sum_{\sigma'}\int_{\epsilon_{\mathrm {Fermi}}}^{\emax}\dd\epsilon'\;n_{\sigma'}(\epsilon',z,t)\times\\
p_{\sigma^{\prime},\sigma}(\epsilon',\epsilon,z,t) & \left(1-e^{-\delta{t}/\tau_{\sigma'}(\epsilon',z,t)}\right) ,
\end{split}
\label{Eq:Sp_def}
\end{equation}
where, in general, $p_{\sigma',\sigma}(\epsilon',\epsilon,z,t)$ is
the probability that an electron at energy level $\epsilon'$, between the Fermi energy $\epsilon_{\mathrm {Fermi}}$ and the energy cut-off 
$\epsilon_{\mathrm {max}}$, and
spin $\sigma'$ will move to energy level $\epsilon$ with spin $\sigma$
in the next time step, $t+\delta{t}$. 
Eq.~(\ref{Eq:Sp_def}) includes both the contributions from the scattered hot electrons, formally treated as newly exited,
and the ones actually excited as the result of the scattering.

In order to include multiple transmissions and reflections in Eq.~\eqref{Eq:n-sol} it is beneficial to calculate the total particle fluxes in a recursive way. Following \citep{Battiato2014_JAP,Lu:PRB2020},   the fluxes are obtained from
\begin{widetext}
\begin{equation}
\begin{split}\PR_\sigma(z_i^+, t)\;= & \sum_{t_{0}=0}^{t}
S_\sigma^{\mathrm {eff}}\left(z_i,t_{0}\right)\psi_\sigma \left(z_i^+,t | z_i,t_{0} \right)\;
+ \sum_{\sigma'} \left[ \TR_
{\sigma, \sigma'}(z_i^-)\,\PR_{\sigma'}\left(z_i^-,t,\right)+
 \RL_{\sigma, \sigma'}(z_i^-)\,\PL_{\sigma'}\left(z_i^-,t,\right) \right] ,
\end{split}
\nonumber
\end{equation}
\begin{equation}
\begin{split}\PL_\sigma(z_i^+, t)\;= & \sum_{t_{0}=0}^{t}
S_\sigma^{\mathrm {eff}}\left(z_{i+1},t_{0}\right)\psi_\sigma \left(z_i^+,t | z_{i+1},t_{0} \right)\;+
 \sum_{\sigma'} \left[ \TL_{\sigma, \sigma'}(z_{i+1}^+)\,\PL_{\sigma'}\left(z_{i+1}^+,t,\right)+
  \RR_{\sigma, \sigma'}(z_{i+1}^+)\,\PR_{\sigma'}\left(z_{i+1}^+,t,\right)\right] ,                 
\end{split}
\label{Eq:flux}
\end{equation}
\end{widetext}
where we have introduced the spin-conserving and spin-flipping
reflection probabilities for left and right moving particles,  
$\RL_{\sigma, \sigma'}$ and $\RR_{\sigma, \sigma'}$.  
Moreover,  in Eq.\ \eqref{Eq:flux} we have introduced the integrated flux, given by
$\psi_\sigma$, for which an analytical expression was given in Ref.~\citealp{Battiato2014_JAP}.
Throughout this paper, we keep the spatial discretization step $\dz = 1\, {\rm nm}$ and
time step $\dt = 1\, {\rm fs}$.

\subsection{Model system}

\label{Sec:Model}

\begin{figure}[t!]
\includegraphics[width=0.95\columnwidth]{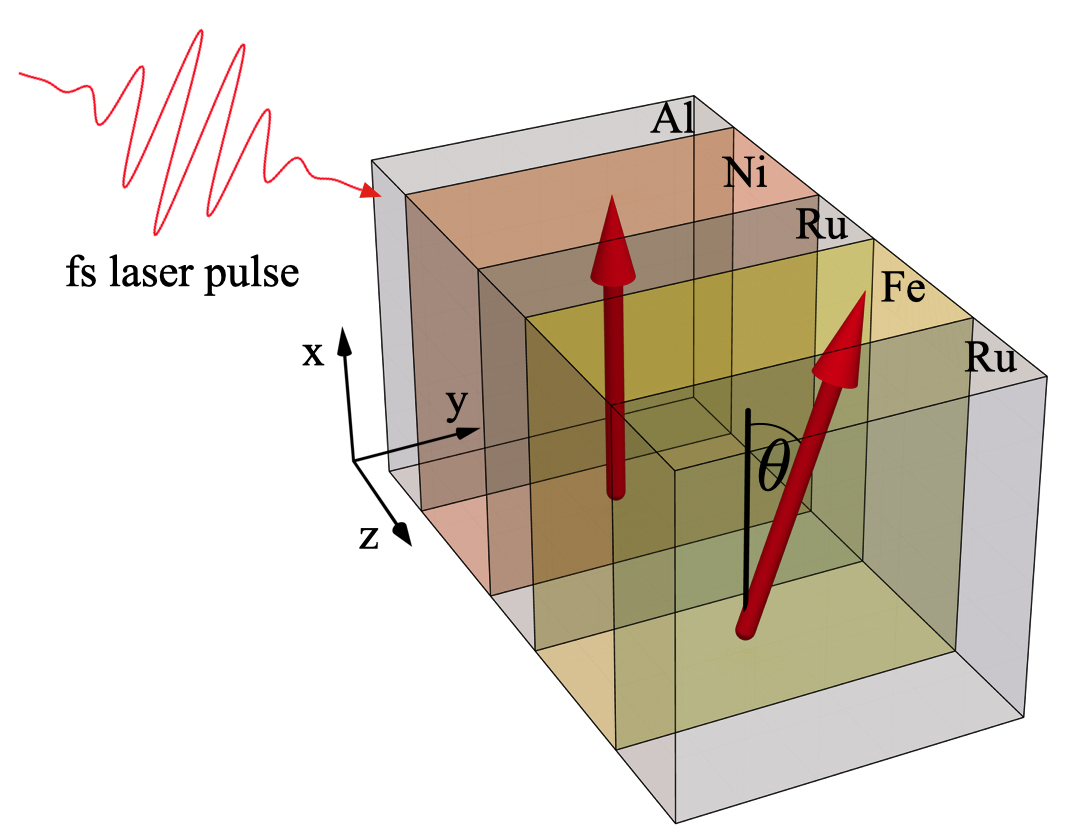} \caption{Model of the studied magnetic multilayer. The laser pulse excites the multilayer on the Al side. The magnetizations of the Ni and and Fe layers are noncollinear, expressed by the angle $\theta$.}
\label{Fig:Model} 
\end{figure}

To investigate the influence of noncollinear magnetizations, we model a magnetic multilayer, Al(3nm)/Ni(5nm)/Ru(2nm)/Fe(4nm)/Ru(5nm),
shown schematically in Fig.~\ref{Fig:Model}~\cite{Rudolf2012,Eschenlohr2017}.
The magnetic layers, Ni and Fe, are separated by a
thin nonmagnetic Ru layer. The magnetization of the Ni magnetic layer is assumed to be aligned with the
$x$ axis, while the magnetization of the Fe
layer can in general be rotated, in the plane of the layer, by an angle $\theta$ from the $x$ axis.

We assume that the laser pulse illuminates the sample from the side of the outer
Al layer and that the hot electrons are excited just in the magnetic
Ni layer. Moreover, for simplicity, the exited electrons are homogeneously
distributed throughout the layer.

A necessary ingredient for solving \cref{Eq:n-sol,Eq:gen_source,Eq:Sp_def,Eq:flux}
are the spin- and energy-dependent transmission probabilities through
the interfaces between the layers. In the next
subsection (Sec.~\ref{Sec:Int_method}) we shall describe the method used for calculating these at
the material-specific level. 

\subsection{The theory of interface scattering}

\label{Sec:Int_method}

The material-specific transmissions through interfaces or multilayers
were obtained using the two-step procedure described in Refs.~\citep{Zwierzycki2008:PSSb,Zwierzycki2005:PRB}.
In the first step the self-consistent potentials for a system of
interest were calculated using the tight-binding linear muffin-tin
orbital method \citep{Andersen:85,Turek:97} (TB-LMTO) in the atomic-sphere
approximation (ASA). The potentials were then used to calculate the
elements of the scattering matrix, {i.e.}\ transmission and reflection
coefficients using the wave function matching method~\citep{Zwierzycki2008:PSSb,Zwierzycki2005:PRB}.
The latter step was performed for energies in the interval $\left[\epsilon_{\rm Fermi},\epsilon_{\rm Fermi}+\unit[1.5]{eV}\right]$ with a step of $\unit[0.125]{eV}$. The length of the interval (\unit[1.5]{eV}) corresponds to the typical energy of the 
photons from the exciting laser in experiments.

Little is known about the crystalline structure and orientation of
the experimental structures \citep{Rudolf2012,Eschenlohr2017}. We
assumed therefore that all the elements grow in their bulk lattice structures,
{i.e.}\ fcc for Al and Ni, bcc for Fe and hcp
for Ru. Furthermore, we chose the $(111)$ orientation for Al, Ni,
and Fe and $(0001)$ for the Ru layer, thus preserving the maximum axial
symmetry. The main difficulty arising when performing the calculations
is the substantial lattice constant mismatch between the different layers. 
In order to take this into account we modeled the interfaces
using lateral supercells of different sizes. In particular, we used
$7\times7|8\times8$ for the interface between Al ($a_{\rm Al}=\unit[4.05]{\text{\AA}}$)
and Ni ($a_{\rm Ni}=\unit[3.52]{\text{\AA}}$), $13\times13|12\times12$
for Ni and Ru ($a_{\rm Ru}=\unit[2.71]{\text{\AA}}$) and $3\times3|2\times2$
for Ru and Fe ($a_{\rm Fe}=\unit[2.87]{\text{\AA}}$). The trilayer structure
was modeled as Ni$(13\times13)|$Ru$(12\times12)|$Fe$(8\times8)$.
The adjustments of the in-plane lattice constants, necessary to obtain
a perfect match between supercells were well below $1\%$. A small
tetragonal distortion was additionally introduced in order to keep
the volume of the unit cells intact. At the interfaces the distance
between the atomic layers was set so as to fulfil the local space-filling
requirement of the ASA approximation, with the radii of the atomic spheres kept at their bulk values.

The ASA potentials for the supercell calculations were obtained using
a simplified procedure described in Ref.~\citealp{Balaz_prb13}. For each interface,
two single unit cell ($1\times 1$) calculations were made with the common lattice
constant set first to the value corresponding to the bulk structure of the material 
to the left and then to the right of the interface. The final interface structure was 
then constructed using the spheres (and the associated potentials) with the correct radius for either side.

With the transmission and/or reflection coefficients calculated, the
energy-dependent Landauer-B\"uttiker (LB) conductances can be calculated as: 
\begin{equation}
G(\sigma,\epsilon)=\frac{e^{2}}{h}\frac{1}{4\pi^{2}}\int_{\rm 2DBZ}\sum_{\nu\mu}\left|t_{\nu\mu}\left(k_{\parallel}\right)\right|^{2}.\label{eq:LB_cond}
\end{equation}
where $t_{\nu\mu}\left(k_{\parallel}\right)$ are the scattering coefficients,
and the $\nu$ and $\mu$ are the indices for the modes at opposite
sides of the interface, characterized by the same lateral component
of the wave vector ($k_{\parallel}$) and the two-dimensional Brillouin
zone (2DBZ) is defined by the matching lateral supercells.  In Eq.~(\ref{eq:LB_cond})
we implicitly assumed the specular (i.e.\ $k_{\parallel}$-preserving)
scattering. However, since the interface between non-lattice-matched
lattices inherently breaks the lateral translational symmetry, a certain
amount of diffusive scattering is still possible even with enforced
perfect match between supercells on both sides of the interface. 
The 2DBZ defined for $N\times N$ lateral supercell is the folded down version 
of the original ($N=1$) 
zone\footnote{Note that the $N=1$ 2DBZs are generally different for the non-latticed matched materials. 
This is the case for our Al/Ni/Ru/Fe multilayer.}. 
Consequently the set of $N^2$ $k_\parallel$-points in the original $N=1$ zone 
is now represented by a single point in the folded-down version. 
All the propagating modes characterized originally by the lateral 
wave vectors belonging to the $N^2$ set are now formally treated as having the same $k_\parallel$ wave vector.
When the supercells on the left and right side of the interface are matched in external dimensions but differ in multiplicity (\emph{i.e.} $N_La_L=N_Ra_R$ but $N_L \neq N_R$, $a_{L(R)}$ being the left- and right-side lattice constants) the lowering of the symmetry may enable matching between the modes originally (\emph{i.e.} in unfolded 2DBZs) belonging to different $k_\parallel$ wave vectors. 

The LB conductance through the interface does not depend on the direction
of scattering. The same however is not true for the Sharvin conductances
\begin{eqnarray}
\overrightarrow{G}_{{\rm Sh}}(\sigma,\epsilon) & = & \frac{e^{2}}{h}\frac{1}{4\pi^{2}}\int_{\rm 2DBZ}\sum_{\mu\mu^{\prime}}\delta_{\mu^{\prime}\mu}=\frac{e^{2}}{h}N_{L} ,\label{eq:GShL}\\
\overleftarrow{G}_{{\rm Sh}}(\sigma,\epsilon) & = & \frac{e^{2}}{h}\frac{1}{4\pi^{2}}\int_{\rm 2DBZ}\sum_{\nu^{\prime}\nu}\delta_{\nu\nu^{\prime}}=\frac{e^{2}}{h}N_{R} , \label{eq:GShR}
\end{eqnarray}
which amount simply to the number of modes (per unit area) in the
left ($N_{L}$) or right ($N_{R}$) lead. Consequently, the averaged
transmission for the left- and right-moving electrons,
\begin{equation}
\TR_{\sigma}=\frac{G(\sigma,\epsilon)}{\GR_{{\rm Sh}}(\sigma,\epsilon)} \quad {\rm and} \quad\TL_{\sigma}=\frac{G(\sigma,\epsilon)}{\GL_{{\rm Sh}}(\sigma,\epsilon)}\,,\label{Eq:trans_single}
\end{equation}
also assume different values. 
The transmissions obtained
both for single interfaces and for the whole Ni/Ru/Fe trilayer will be discussed in Sec.~\ref{Sec:Int_results}. 

\section{The transport through interfaces}
\label{Sec:Int_results}

\subsection{Single interfaces}

\begin{figure*}[htp]
    \centering 
    \includegraphics[width=1\textwidth]{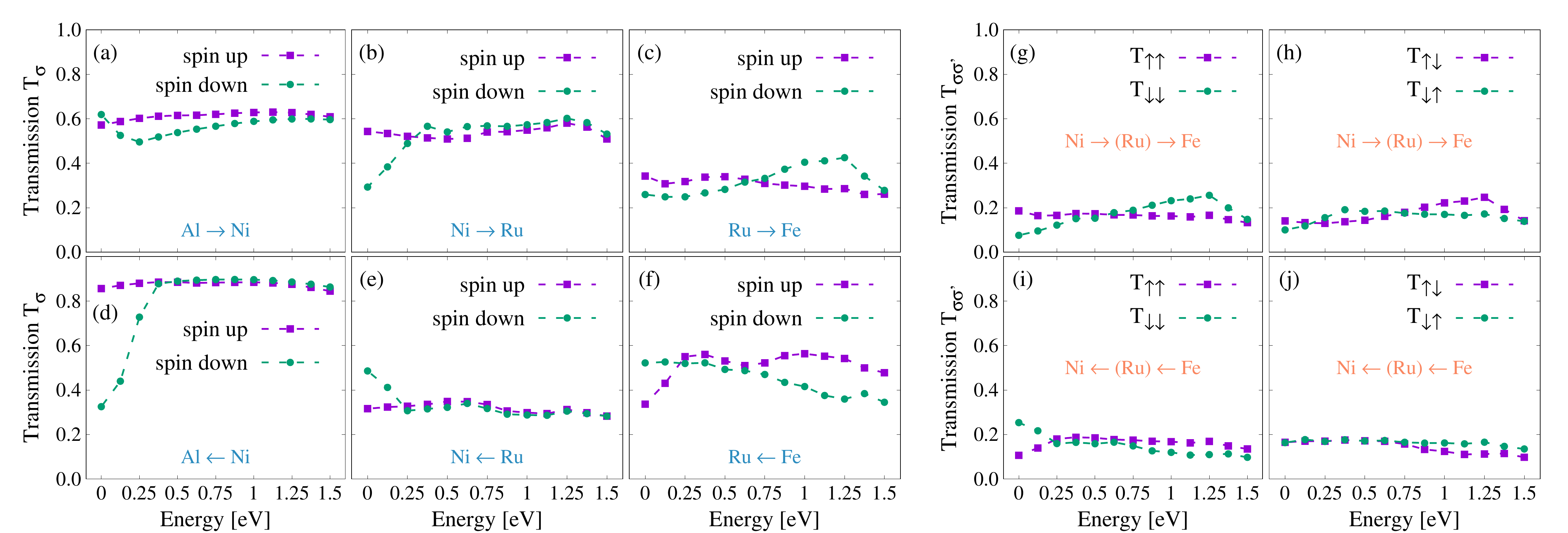} 
    \caption{\label{Fig:transmissions} 
    The spin-dependent transmissions of hot electrons as a function of energy, calculated using {\em ab
    initio} methods for each single interface in the Al/Ni/Ru/Fe/Ru multilayer.
    The top row shows transmissions for electrons moving from the left
    to the right, while the bottom row shows transmissions for electrons moving from the right
    to the left.\label{Fig:T_int}}
\end{figure*}

Figure~\ref{Fig:transmissions} shows the calculated 
transmissions for hot electrons in the spin-up and spin-down channels
as a function of energy for each single interface in the studied multilayer,
namely, Al/Ni, Ni/Ru, and Ru/Fe. We show separately the transmissions
for electrons moving from the left to the right (top row) and electrons
moving from the right to the left (bottom row). In all three cases,
one can observe a significant nonlinear dependence of transmission
on energy. Moreover, a strong asymmetry between spin-up and spin-down
channels is obvious for all three interfaces for some energies. 

The results for Al/Ni interface [Figs.~\ref{Fig:T_int}(a) and (b)]
are in agreement with those published recently~\citep{Lu:PRB2020}.
The small differences are easily explainable by the different choices
of interface geometry, {i.e.}\ crystallographic orientations of the
respective metals, between Ref.~\citealp{Lu:PRB2020} and the present paper.

\subsection{Central nonmagnetic layer}

The key part of our model is the treatment of the central nonmagnetic
layer, which separates two magnetic layers, Ni and Fe. If the magnetizations
in the two magnetic layers are collinear, {i.e.}\ parallel or
antiparallel, the electronic and spin transport can be described by the
two sets of independent equations for the spin-up and spin-down channel,
making use of the single-interface transmission parameters shown
in Fig.~\ref{Fig:transmissions}. This case
has been studied experimentally and modeled using the superdiffusive
spin-dependent transport theory in Ref.~\citealp{Rudolf2012}, although without the \emph{ab initio} calculated scattering coefficients.

In the case of a noncollinear magnetic configuration, the spin channels are 
mixed.\footnote{Note that we do not take into account explicit spin-flip scattering at the interfaces.}
Consequently, the spin currents in the Ru spacer now possess the component perpendicular to the magnetizations of magnetic layers. This in turn leads to spin torques acting on the moments of both Ni an Fe layers. 
In our model, we assume that the central nonmagnetic layer
is thin enough ($\sim1-2\,{\rm nm}$, as in experiment \cite{Rudolf2012}) so that the interface scattering
dominates over the contribution from the bulk of the layer.

In order to asses the importance of the multiple internal reflections
in the central layer we shall first consider a simplified model
where these are ignored and only the first order contribution to the
total transmission probability is taken into account.

For the electrons moving to the right we define the diagonal transmission
through the single interface as
\begin{equation}
    \TRmat^{(i)}=\begin{pmatrix}
        \TRu^{(i)} & 0\\
        0 & \TRd^{(i)}
        \end{pmatrix}
\label{Eq:trans1}
\end{equation}
with $i=1,2$ corresponding to Ni/Ru and Ru/Fe interface, respectively.

Ignoring the internal reflections in the central layer we can express
the total transmission of the spacer and internal interfaces as
\begin{equation}
\TRmat=\TRmat^{(1)}\,\Rmat(\theta)\,\TRmat^{(2)}\,,\label{Eq:trans_total_def}
\end{equation}
where $\underline{R}(\theta)$ is the spin rotation matrix due to
noncollinear magnetic configurations of Ni and Fe, defined as 
\begin{equation}
\Rmat(\theta)=\begin{pmatrix}\cos^{2}(\theta/2) & \sin^{2}(\theta/2)\\
\sin^{2}(\theta/2) & \cos^{2}(\theta/2)
\end{pmatrix},\label{Eq:rot}
\end{equation}
with $\theta$ being the angle between their magnetizations. The transmissions
introduced in Eqs.\ (\ref{Eq:trans_single}) and (\ref{Eq:trans1}) are
\emph{probabilities} therefore the formula for total transmission
explicitly assumes the loss of coherence between the two interfaces.

Finally, we obtain 
\begin{equation}
\TRmat=\begin{pmatrix}\TRuu\,\cos^{2}(\theta/2) & \TRud\,\sin^{2}(\theta/2)\\
\TRdu\,\sin^{2}(\theta/2) & \TRdd\,\cos^{2}(\theta/2)
\end{pmatrix} ,\label{Eq:trans_total}
\end{equation}
where 
\begin{subequations}
\begin{align}
\TRuu & =\TRu^{(1)}\TRu^{(2)}\,,\\
\TRud & =\TRu^{(1)}\TRd^{(2)}\,,\\
\TRdu & =\TRd^{(1)}\TRu^{(2)}\,,\\
\TRdd & =\TRd^{(1)}\TRd^{(2)}\,.
\end{align}
\label{Eqs:trans_coefs} 
\end{subequations}

Assuming electrons moving from the left to the right, we can express
the currents in the right magnetic layer as
\begin{equation}
\begin{pmatrix}\jRu^{{\rm R}}\\
\jRd^{{\rm R}}
\end{pmatrix}=\TRmat\;\begin{pmatrix}\jRu^{{\rm L}}\\
\jRd^{{\rm L}}
\end{pmatrix},
\label{Eq:j_RL}
\end{equation}
where $j^{\rm L}_{\sigma}$ and $j^{\rm L}_{\sigma}$ are the majority and minority 
spin-channel currents in the left and right magnetic layer, respectively. Note that the $\sigma$ in the present context should be read as relative to the local magnetization, {i.e.}\ it denotes rather the local majority or minority character than the global orientation of electron spin.

Working out Eq.\ (\ref{Eq:j_RL}) we can write 
\begin{subequations}
\begin{align}
\jRu^{{\rm R}} & =\TRuu\,\cos^{2}(\theta/2)\,\jRu^{{\rm L}}+\TRud\,\sin^{2}(\theta/2)\,\jRd^{{\rm L}}\,,\\
\jRd^{{\rm R}} & =\TRdd\,\sin^{2}(\theta/2)\,\jRu^{{\rm L}}+\TRdd\,\cos^{2}(\theta/2)\,\jRd^{{\rm L}}\,.
\end{align}
\end{subequations}
The spin reflections and transmissions have to obey 
\begin{subequations}
\begin{align}
\RRu+\TRuu\,\cos^{2}(\theta/2)+\TRdu\,\sin^{2}(\theta/2) & =1\,,\\
\RRd+\TRdd\,\cos^{2}(\theta/2)+\TRud\,\sin^{2}(\theta/2) & =1\,,
\end{align}
\label{Eqs:reflections} 
\end{subequations}
\noindent
where $\RRu$ and $\RRd$ are reflections for electrons without change
of their spin. Spin-flip reflections from the interfaces are not taken
into account. Thus, the reflected currents on the left-hand side
(moving to the left) read 
\begin{subequations}
\begin{align}
\jLu^{{\rm L}} & =\left[1-T_{\uparrow\uparrow}\,\cos^{2}(\theta/2)-T_{\downarrow\uparrow}\,\sin^{2}(\theta/2)\right]\,\jRu^{{\rm L}}\,,\\
\jLd^{{\rm L}} & =\left[1-T_{\downarrow\downarrow}\,\cos^{2}(\theta/2)-T_{\uparrow\downarrow}\,\sin^{2}(\theta/2)\right]\, \jRd^{{\rm L}}\,.
\end{align}
\label{Eqs:curr_reflected} 
\end{subequations}
\noindent
Analogically, by interchanging the electrons' directions, one can rewrite
Eqs.~(\ref{Eq:trans1})\,--\,(\ref{Eqs:curr_reflected}) for the
currents moving in the opposite direction.

\subsection{Multiple scattering}

The full treatment of the hot electron transiting through the Ni/Ru/Fe
trilayer requires taking into account the multiple internal reflections
within the Ru spacer. The total transmission matrix for electrons
moving from the left to the right is defined as 
\begin{equation}
\TRmat=\TRmat^{(2)}\Rmat(\theta)\left[\sum_{n=0}^{\infty}\left(\RLmat^{(1)}\Rmat(\theta)\RRmat^{(2)}\Rmat(\theta)\right)^{n}\right]\TRmat^{(1)}\,,\label{Eq:TtotLR}
\end{equation}
where 
\begin{equation}
\RLmat^{(1)}=\Imat-\TLmat^{(1)}=\begin{pmatrix}\RLu^{(1)} & 0\\
0 & \RLd^{(1)}
\end{pmatrix}
\end{equation}
is the reflection matrix of the first interface (Ni/Ru) for electrons
moving to the left, where $\Imat$ is $2\times2$ unit matrix, and
\begin{equation}
\RRmat^{(2)}=\Imat-\TRmat^{(2)}=\begin{pmatrix}\RRu^{(2)} & 0\\
0 & \RRd^{(2)}
\end{pmatrix}
\end{equation}
is the reflection matrix of the second interface (Ru/Fe) for electrons
moving to the right. It can be easily seen that the simplified transmission
given by Eq.~(\ref{Eq:trans_total_def}) is the first term of the series
which constitutes the full transmission in Eq.\ \eqref{Eq:TtotLR}. Even
though formula \eqref{Eq:TtotLR} now incorporates the possibility
of multiple internal reflections it still assumes, just like the simplified
version, the lack of coherence. 

Similarly, we can define the total transmission matrix for electrons
moving from the right to the left as 
\begin{equation}
\TLmat=\TLmat^{(1)}\Rmat(\theta)\left[\sum_{n=0}^{\infty}\left(\RRmat^{(2)}\Rmat(\theta)\RLmat^{(1)}\Rmat(\theta)\right)^{n}\right]\TLmat^{(2)} .\label{Eq:TtotRL}
\end{equation}

\begin{figure*}[htb!]
\centering \includegraphics[width=1\textwidth]{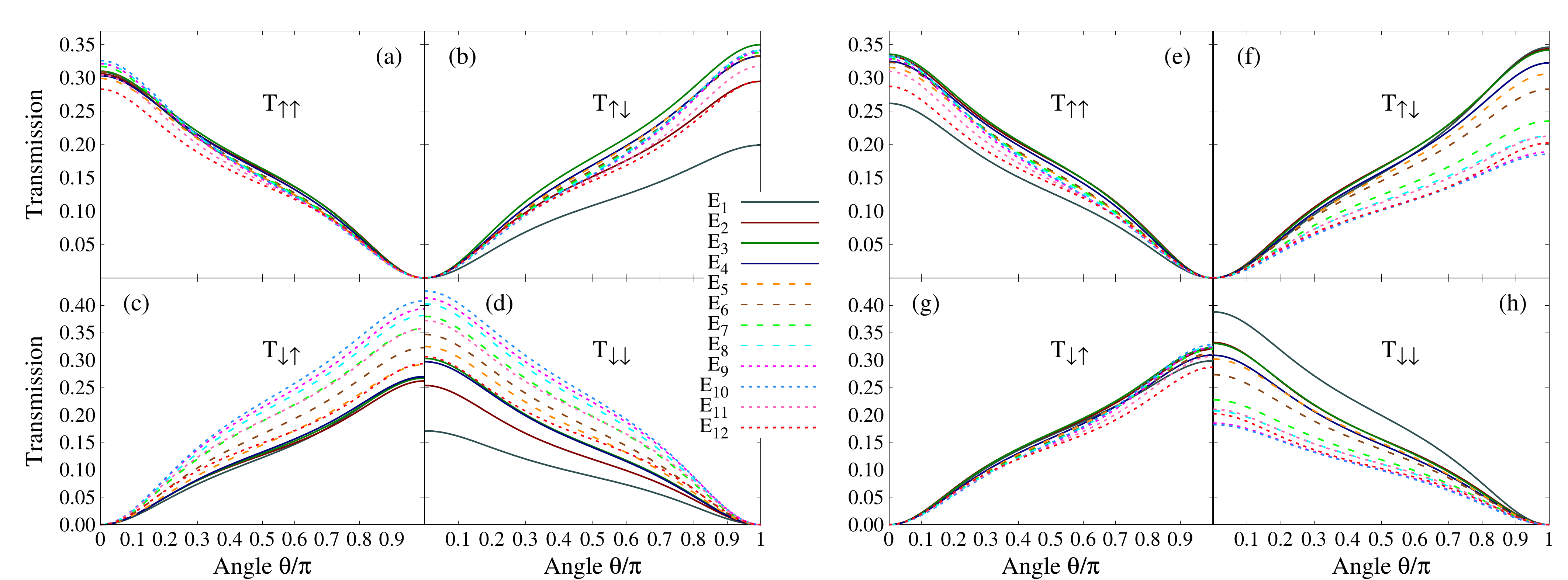}
\caption{\label{Fig:trans_angular} Angular dependence of the total transmission
through the nonmagnetic spacer assuming multiple scattering in the
central Ru layer calculated for 12 energy levels above the Fermi energy,
$E_{n}=(0.125\,{\rm eV})n$. Figures (a) -- (d) show transmissions
for electrons moving from the left to the right, whiles figures (e)
-- (h) correspond to transmissions of electrons moving from the right
to the left. The diagonal elements $T_{\uparrow\uparrow}$, shown
in (a) and (e), and $T_{\downarrow\downarrow}$, shown in (d) and
(h), describe spin-conserving transmissions for the spin-up and spin-down
channel, respectively. The off-diagonal elements, $T_{\uparrow\downarrow}$,
shown in (b) and (f), and $T_{\downarrow\uparrow}$, shown in (c)
and (g), denote the spin-flip transmissions of electrons changing their
spin from down to up and up to down, respectively.}
\end{figure*}

In the parallel magnetic configuration, for $\theta=0$, the above equations lead to 
\begin{subequations}
\begin{align}
\TR_{\sigma} & =\TR_{\sigma}^{(1)}\,\TR_{\sigma}^{(2)}\left[\sum_{n=0}^{\infty}\left(\RL_{\sigma}^{(1)}\,\RR_{\sigma}^{(2)}\right)^{n}\right],\\
\TL_{\sigma} & =\TL_{\sigma}^{(1)}\,\TL_{\sigma}^{(2)}\left[\sum_{n=0}^{\infty}\left(\RL_{\sigma}^{(1)}\,\RR_{\sigma}^{(2)}\right)^{n}\right],
\end{align}
\end{subequations}
for both $\sigma\in\{\uparrow,\downarrow\}$, which has as solution
\begin{subequations}
\begin{align}
\TR_{\sigma} & =\frac{\TR_{\sigma}^{(1)}\,\TR_{\sigma}^{(2)}}{1-\RL_{\sigma}^{(1)}\,\RR_{\sigma}^{(2)}}\,,\\
\TL_{\sigma} & =\frac{\TL_{\sigma}^{(1)}\,\TL_{\sigma}^{(2)}}{1-\RL_{\sigma}^{(1)}\,\RR_{\sigma}^{(2)}}\,.
\end{align}
\label{Eqs:Ttot_configP} 
\end{subequations}
\noindent
Thus, considering internal multiple reflections in the nonmagnetic
spacer might substantially increase the total transmissions. Moreover,
the angular dependence of the transmissions might depart from the
simple sine-like dependence.
Fig.~\ref{Fig:trans_angular} shows the angular dependence of the
transmission probabilities
calculated for different energies, $E_{n}=(0.125\,{\rm eV})n$
with $n=1,2,\dots,12$, above the Fermi level. 

Panels \ref{Fig:trans_angular}(a) -- \ref{Fig:trans_angular}(d) show transmissions for electrons moving from the
left to the right, while panels \ref{Fig:trans_angular}(e) -- \ref{Fig:trans_angular}(h) contain transmissions
for electrons moving from the right to the left. In both cases we
show the diagonal elements of the transmission matrices, $T_{\uparrow\uparrow}$
and $T_{\downarrow\downarrow}$, which correspond to spin-conserving
transmissions through the spacer for the spin-up and spin-down channel,
respectively. The off-diagonal elements of the transmission matrix,
$T_{\uparrow\downarrow}$ and $T_{\downarrow\uparrow}$, correspond
to the spin-flip transmissions, which mix the spin channels. Obviously,
the spin-conserving transmissions reach their maximum for $\theta=0$,
i.e.\ in the parallel magnetic configuration, when the spin-flip terms
reach zero. Oppositely, for $\theta=\pi$, i.e.\ in the antiparallel
magnetic configuration, the electrons passing the nonmagnetic spacer
are injected into the opposite spin channel. Thus, the spin-flip transmissions
are maximal, while the spin-conserving transmissions disappear.

The above described model for the transmissions of electrons through
the noncollinear system combines results of {\em ab initio} calculations
with the classical  formula for non-coherent multiple reflections.
To test the model we shall compare the transmissions obtained from \eqref{Eq:trans_total_def}
and \eqref{Eq:TtotLR} with the results of explicit calculations, 
by using the method outlined in Sec.\,\ref{Sec:Int_method}, of spin-dependent
transmission through the Ni/Ru(2nm)/Fe trilayer, where Ni and Fe form
the semi-infinite electrodes. 

\begin{figure}[htp!]
\centering \includegraphics[width=1.0\columnwidth]{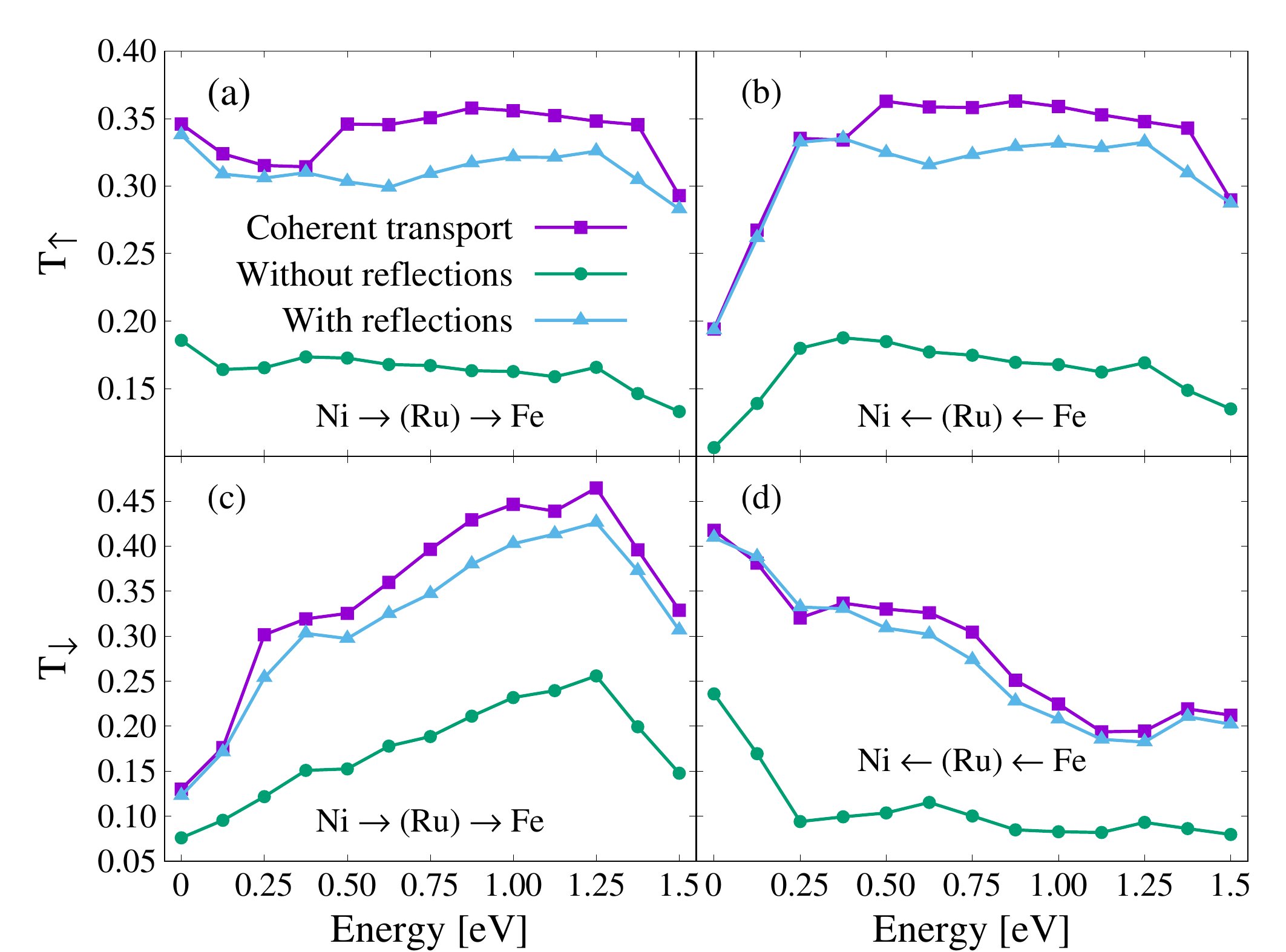} 
\caption{\label{Fig:transP_compare} Comparison of the energy dependence of electron
transmissions through the nonmagnetic spacer of Ni/Ru/Fe, calculated using
three different models: quantum coherent transport (squares), Eq.~(\ref{Eq:trans_total_def})
without considering multiple internal reflections in the Ru layer
(circles), and Eq.~(\ref{Eq:TtotLR}) considering internal multiple
reflections of electrons on the Ru interfaces (triangels). Panels (a) and (b)
show transmissions of electrons in the spin-up channel, while panels
(c) and (d) show transmissions of electrons in the spin-down channel.
In panels (a) and (c) we plot transmissions of electrons moving from
the left to the right (Ni$\rightarrow$Fe), in panels (b) and (d)
transmissions for electrons moving from the right to the left (Ni$\leftarrow$Fe)
are plotted.}
\end{figure}

A comparison of all three models of the
Ni/Ru/Fe transmissions is shown in Fig.~\ref{Fig:transP_compare}.
Comparing the semiclassical transport model model with and without internal reflections, one can
notice a significant increase of transmissions when reflections are
included. Furthermore, it can be seen that the semiclassical approach
including internal multiple reflections is in a reasonably good agreement
with the fully quantum mechanical approach. Therefore, in the next
 sections we shall adapt the semiclassical approach including multiple
internal reflections to model the spin-dependent transport through
the thin nonmagnetic spacer.

Having established the way to calculate transmissions for a noncollinear
system, we can now proceed with the solution of equation (\ref{Eq:n-sol}).
From this equation we obtain the time dependence of the spin density as well as
spin fluxes for each energy and spin channel in each point of the
discretized trilayer. This allows us to inspect the laser-induced
demagnetization in the magnetic layers and to define the spin-transfer
torques acting on the magnetizations. The results will be discussed
in the following sections.

\section{Demagnetization}
\label{Sec:Demag}

In this section we focus on the effect of laser-induced ultrafast
demagnetization in the studied trilayer. We start with the collinear
magnetic configurations. Since this case can be described within the
original model of superdiffusive hot electrons transport \citep{Battiato2014_JAP},
the comparison of the results with the model used in this paper will
serve as a test of the validity of our approach, in particular the
assumption of the negligible spacer thickness. In the later part of
this section, we will focus on the angular dependence of the demagnetization
in both magnetic layers.

In our calculations we assumed that the original nonthermal excitation
is caused by the Gaussian laser pulse with the halfwidth $t_{p}=\unit[35]{fs}$.
In the experimental study by Rudolf {\em et al.}, 
Al and Ni layers together absorb more than $58\%$ of the incident
light, which was about $2.5$ times as much as the Fe layer~\cite{Rudolf2012}.
Thus, for simplicity, we assumed that the excited electrons are homogeneously
distributed throughout the Ni layer and nowhere else.
The energy distribution
of the excited population was set to be constant over the interval $\left[\epsilon_{\rm Fermi},\epsilon_{\rm Fermi}+\unit[1.5]{eV}\right]$.
In the calculations the energy dependence over this range was discretized
on a grid with $\Delta\epsilon=\unit[0.125]{eV}$, known from previous
experience to be sufficiently precise. 
We assumed that the areal electron density excited by the laser pulse 
is 0.1 electron per energy/spin level in each computational cell of the Ni layer.

\subsection{Collinear magnetic configurations}
\label{SSec:DemagCol}

\begin{figure}[!htp]
\centering \includegraphics[width=0.9\columnwidth]{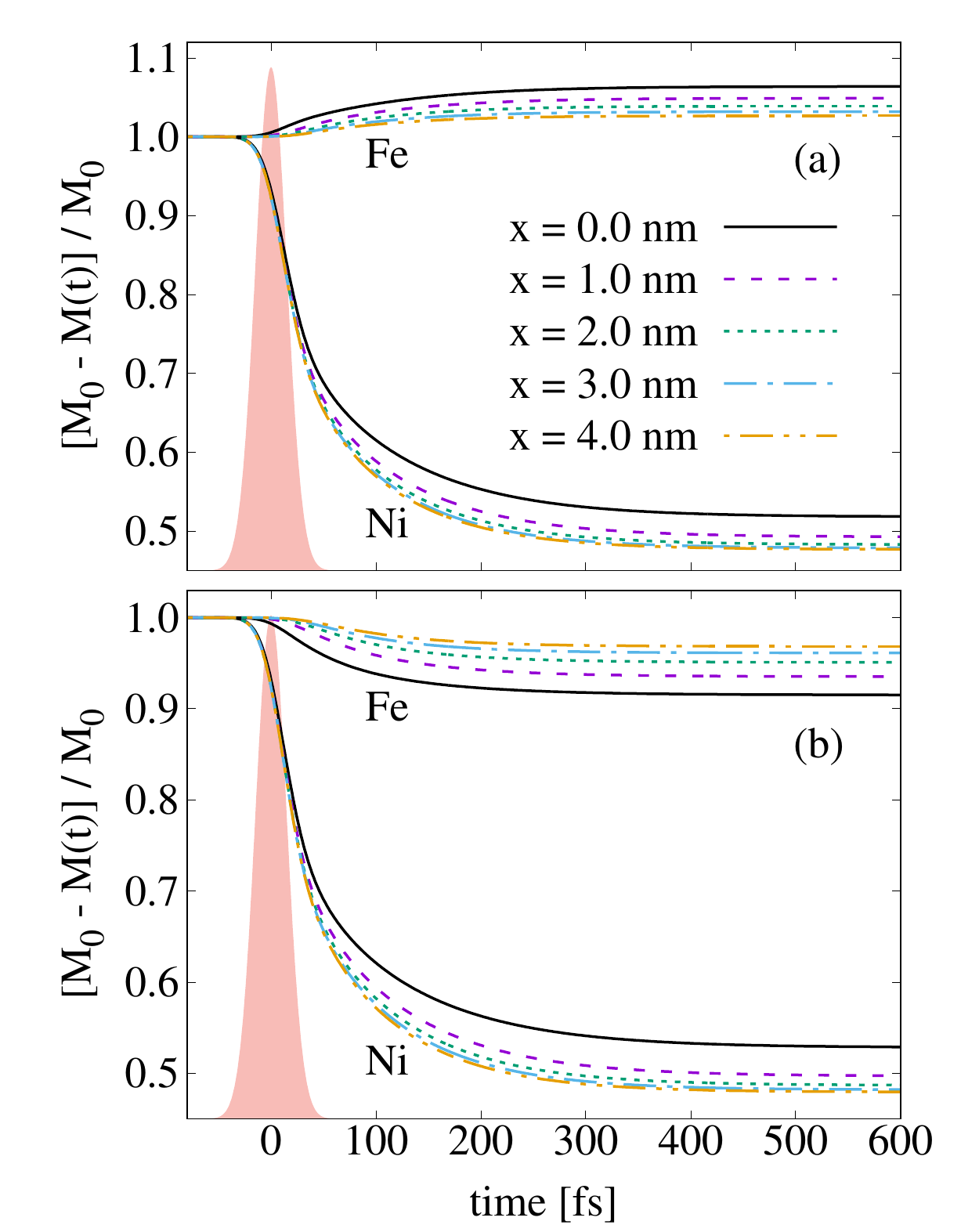}
\caption{\label{Fig:demag_col} Time dependence of the magnetization of the Fe and
Ni layer in the Al(3)/Ni(5)/Ru($x$)/Fe(5)/Ru multilayer due to the laser
pulse, (a) in the parallel magnetic configuration, and (b) in the antiparallel
magnetic configuration. The time dependence of the laser fluence is
given by the red filled area.}
\end{figure}

Fig.~\ref{Fig:demag_col} shows the time evolution of the magnetization in the
Ni and Fe magnetic layer in the studied multilayer due to the laser pulse, calculated for the
collinear magnetic configurations, parallel (P) and antiparallel (AP). The results for varying thickness 
of the Ru spacer, given by $x$, are plotted using the lines of different type and color. The $x=0.0\;\rm{nm}$
case was calculated using the multiple-reflections formulas (\ref{Eq:TtotLR})
and (\ref{Eq:TtotRL}) and the values of interface transmissions from Figs.~\ref{Fig:transmissions}(a)\,--\,(f).
The finite thickness cases were calculated using the standard, collinear, version of the superdiffusive transport 
model of Ref.~\citealp{Battiato2014_JAP} which is applicable in this case. Note that the multiple reflections within the spacer are naturally included in this version of the formalism. We further note that (diffusive) remagnetization processes \cite{Yastremsky2014} are not included in the simulations.

Comparing the results for $x=0$ with the finite-width spacer model,
one can see a qualitative agreement of both models. Importantly, in
accord with the previous calculations and experiments~\citep{Rudolf2012}
we observe an \textit{enhancement} of the Fe magnetization in the P configuration.
In the AP configuration the magnetizations decrease in both layers. 
As the spacer width increases, the demagnetization of the Ni layer
increases. This is because in the model with $x=0$ electrons moving 
towards the Fe layer have to overcome the total reflectivity of both iterfaces
(Ni/Ru and Ru/Fe). In the model with finite thickness the electrons can
flow into the Ru layer. Since the transmission of Ni/Ru interface 
[Fig.~\ref{Fig:T_int}(b)] is much higher than the one of the effective
Ni/Fe interface [Fig.~\ref{Fig:T_int}(g)], the demagnetization is higher when 
$x > 0$. Moreover, with increasing $x$, the demagnetization approaches a constant value,
which is a sign of a finite spin diffusion length in the Ru layer.
Similarly, in the P configuration the laser-induced magnetization enhancement in the Fe layer
becomes smaller as $x$ increases, because some electrons relax in the Ru layer, or
are reflected from the Ru/Fe interface. 
Oppositely, in the AP configuration one can observe a smaller demagnetization in the Fe layer
for the same reason.

We can conclude that the zero-width model does reproduce correctly the trends observed in calculations with a finite spacer thickness. What's more the numerical results do not deviate too strongly from those corresponding to experimental thicknesses {i.e.}\ $x\sim1-2~{\rm nm}$.

\subsection{Noncollinear magnetic configurations}
\label{SSec:DemagNoncol}

\begin{figure}[htp!]
\centering \includegraphics[width=0.9\columnwidth]{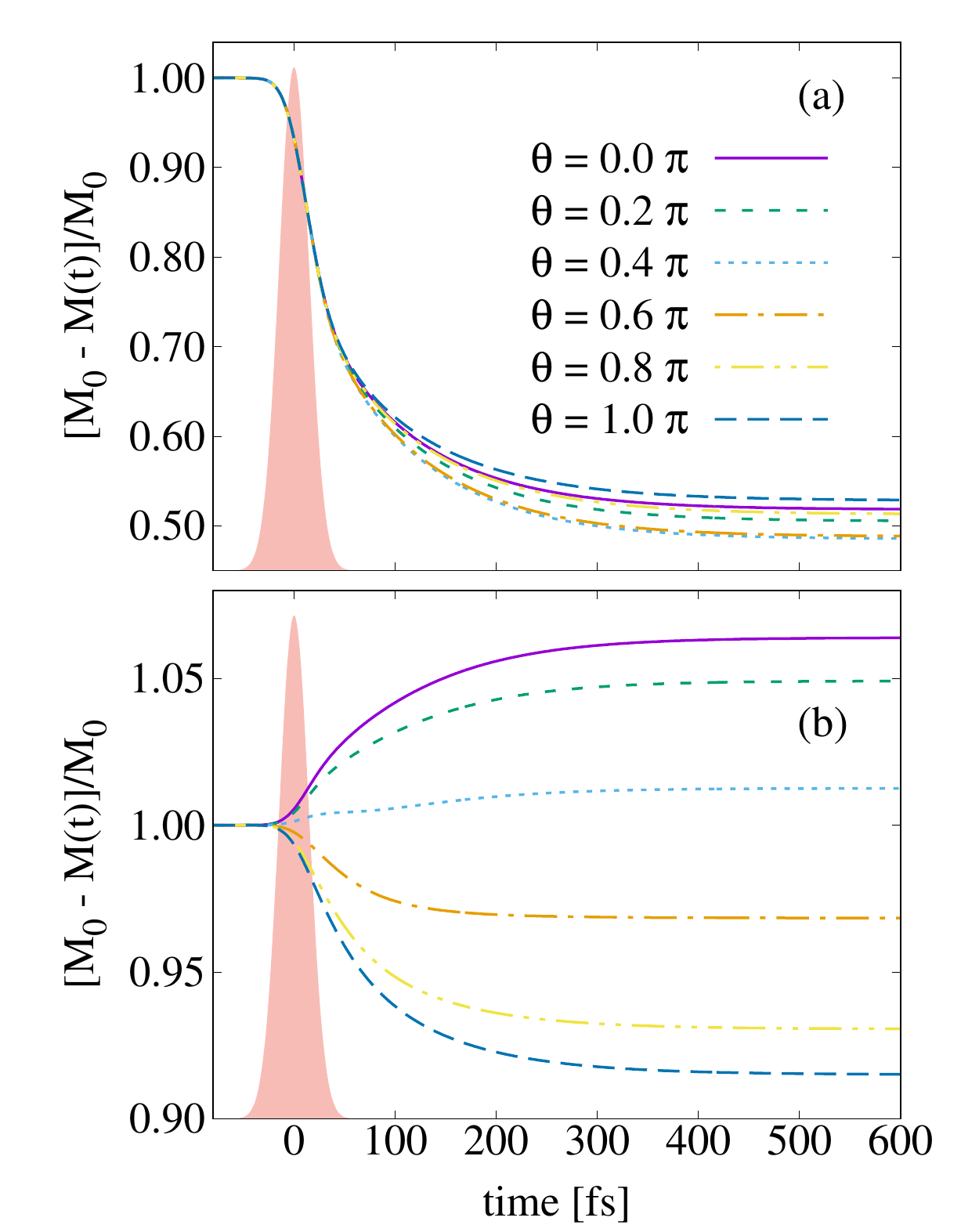}
\caption{\label{Fig:demag_noncol} Time dependence of the magnetization in (a) the nickel and 
(b) the iron layer caused by the femtosecond laser
pulse, calculated in the collinear and noncollinear magnetic configurations
given by the angle $\theta$. The time dependence of the laser fluence
is given by the red filled area.}
\end{figure}

\begin{figure}[htp!]
\centering \includegraphics[width=0.9\columnwidth]{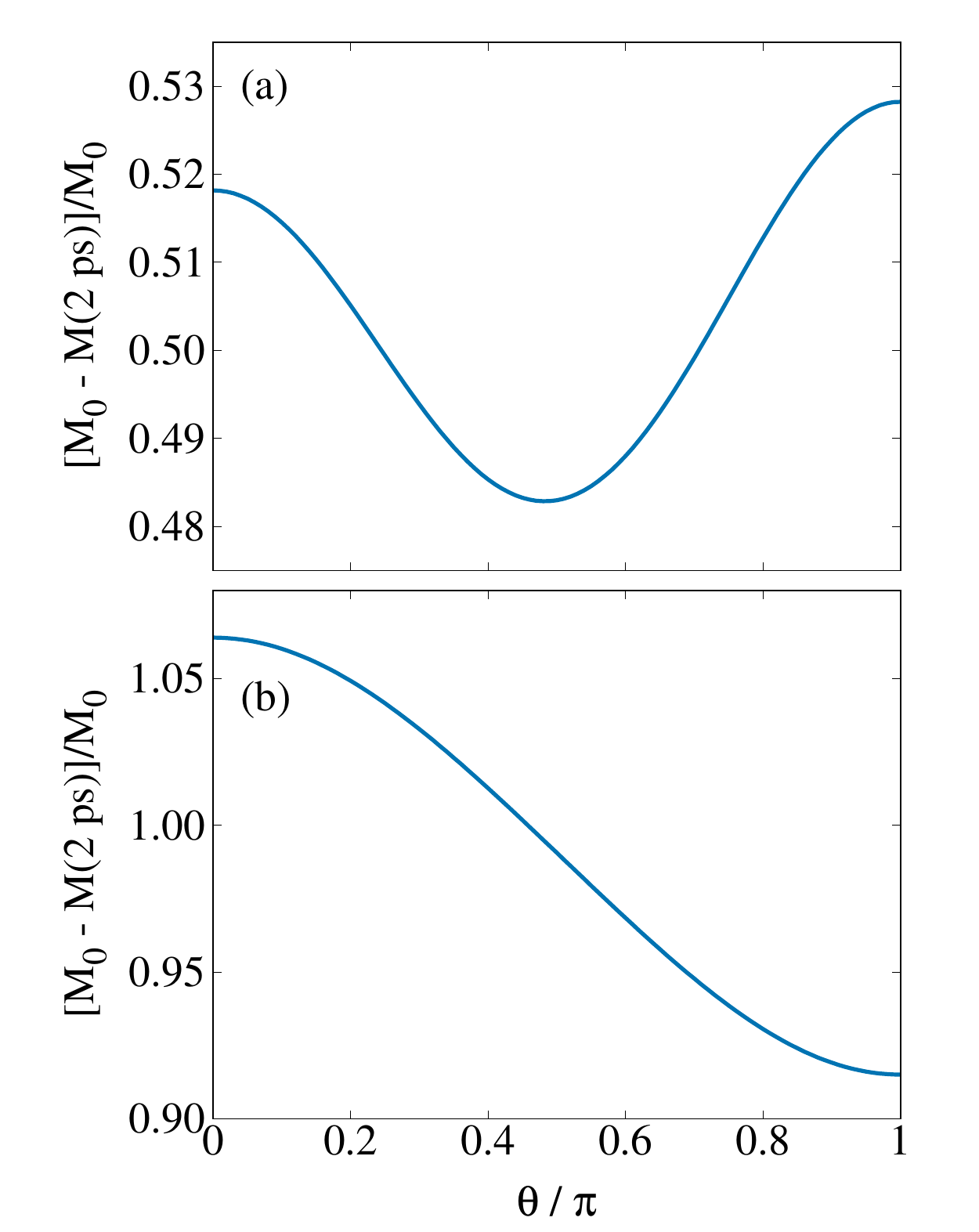}
\caption{\label{Fig:demag_angular} 
The demagnetization of (a) 
the nickel layer and (b) the iron layer as a function of angle $\theta$ between the magnetizations, 
calculated at 2 ps after irradiation with a femtosecond laser pulse.}
\end{figure}

Next we shall focus on the noncollinear magnetic configurations. Fig.~\ref{Fig:demag_noncol}
shows the demagnetization of Ni and Fe layers calculated for different
angles between their magnetization directions. The calculations were done in
the zero-width spacer limit. First, we can notice, that the
magnetic configuration of the Fe layer,
given by angle $\theta$, has almost no influence on the demagnetization
of the Ni layer. This is because Ni is demagnetized directly by the
laser pulse. The electrons reflected from the spacer or secondary
electrons generated in the multilayer have just a minor effect on the
total demagnetization of Ni. In contrast, the magnetization of the Fe layer
is strongly affected by the magnetic configuration. As the angle $\theta$
changes from $0$ to $\pi$, we can see a direct crossover from magnetization
enhancement in the Fe layer, to magnetization reduction. In Fig.~\ref{Fig:demag_angular}
we plot the total angular dependence of the demagnetization as a function
of $\theta$. To this end, we calculated the magnetization of Ni and Fe
layers after $2\,{\rm ps}$, when all the spin currents in the system have
become zero. While the Fe magnetization monotonously decreases
with $\theta$, the magnetization in the Ni layer shows a global minimum,
which, although shallow, corresponds to a maximum demagnetization for a certain noncollinear
configuration as shown in Fig.~\ref{Fig:demag_angular}.

\section{Spin-transfer torque}

\label{Sec:STT}

Lastly, we inspect the spin-transfer torque acting on both magnetic
layers. To define the STT action in the Fe magnetic layer, we need
to calculate the spin current of the electrons moving from the left to
the right at the left- and right-hand side of the Ru spacer. In other
words, we need to calculate the spin current heading from Ni to
the Ru spacer and the one incoming into the Fe layer. The spin current
is calculated as~\citep{r_02_StilesAnatomy}
\begin{equation}
\overrightarrow{\bm{J}}_{{\rm s}}(t)=\frac{\hbar}{2}\sum_{i=1}^{N_{\epsilon}}\left[\overrightarrow{j_{\uparrow}}(\epsilon_{i},t)-\overrightarrow{j_{\downarrow}}(\epsilon_{i},t)\right]\hat{\bm{m}}_{{\rm Ni}}\,,\label{eq:spin_curr_Ni}
\end{equation}
where $N_{\epsilon}=12$ is the number of the hot electrons energy
levels used in the simulation, and $\hat{\bm{m}}_{{\rm Ni}}$ is a
unit vector along the Ni magnetization. Likewise, the spin current on
the right-hand side is given by 
\begin{equation}
\overrightarrow{{\bm{J}}'}_{{\rm s}}(t)=\frac{\hbar}{2}\sum_{i=1}^{N_{\epsilon}}\left[\overrightarrow{j'_{\uparrow}}(\epsilon_{i},t)-\overrightarrow{j'_{\downarrow}}(\epsilon_{i},t)\right]\hat{\bm{m}}_{{\rm Fe}}\,,\label{eq:spin_curr_Fe}
\end{equation}
where $\hat{\bm{m}}_{{\rm Fe}}$ is a unit vector along the Fe magnetization.
Note that the form of Eqs.~(\ref{eq:spin_curr_Ni}) and (\ref{eq:spin_curr_Fe})
describe only the longitudinal spin currents with implicit assumption
that the transversal components are entirely absorbed in the interface
zone (cf.\ \cite{Ghosh2012:PRL}). The total spin-transfer
torque acting on the Fe magnetization is thus~\citep{Barnas2005}
\begin{equation}
{\bm{\tau}}_{{\rm Fe}}=-\left[{\bm{J}}_{{\rm s}}(t)-{\bm{J}}'_{{\rm s}}(t)\right].
\end{equation}
The STT ${\bm{\tau}}_{{\rm Fe}}$ cannot influence the magnitude of the magnetization.
Thus, we are interested only in the STT component, which lies in the
plane defined by the vectors $\hat{\bm{m}}_{{\rm Ni}}$ and $\hat{\bm{m}}_{{\rm Fe}}$,
which can change the 
angles between the two magnetizations,
\begin{equation}
\tau_{\theta\,{\rm Fe}}(t)={\bm{\tau}}_{{\rm Fe}}\cdot\hat{\bm{e}}_{\theta} ,
\end{equation}
where $\hat{\bm{e}}_{\theta}=(\cos\theta,0,-\sin\theta)$.
Analogically, one can define the STT acting on the Ni magnetic layer, taking
into account hot electrons moving from the Fe magnetic layer through
the Ru spacer.

\begin{figure}[htp!] 
\centering \includegraphics[width=0.9\columnwidth]{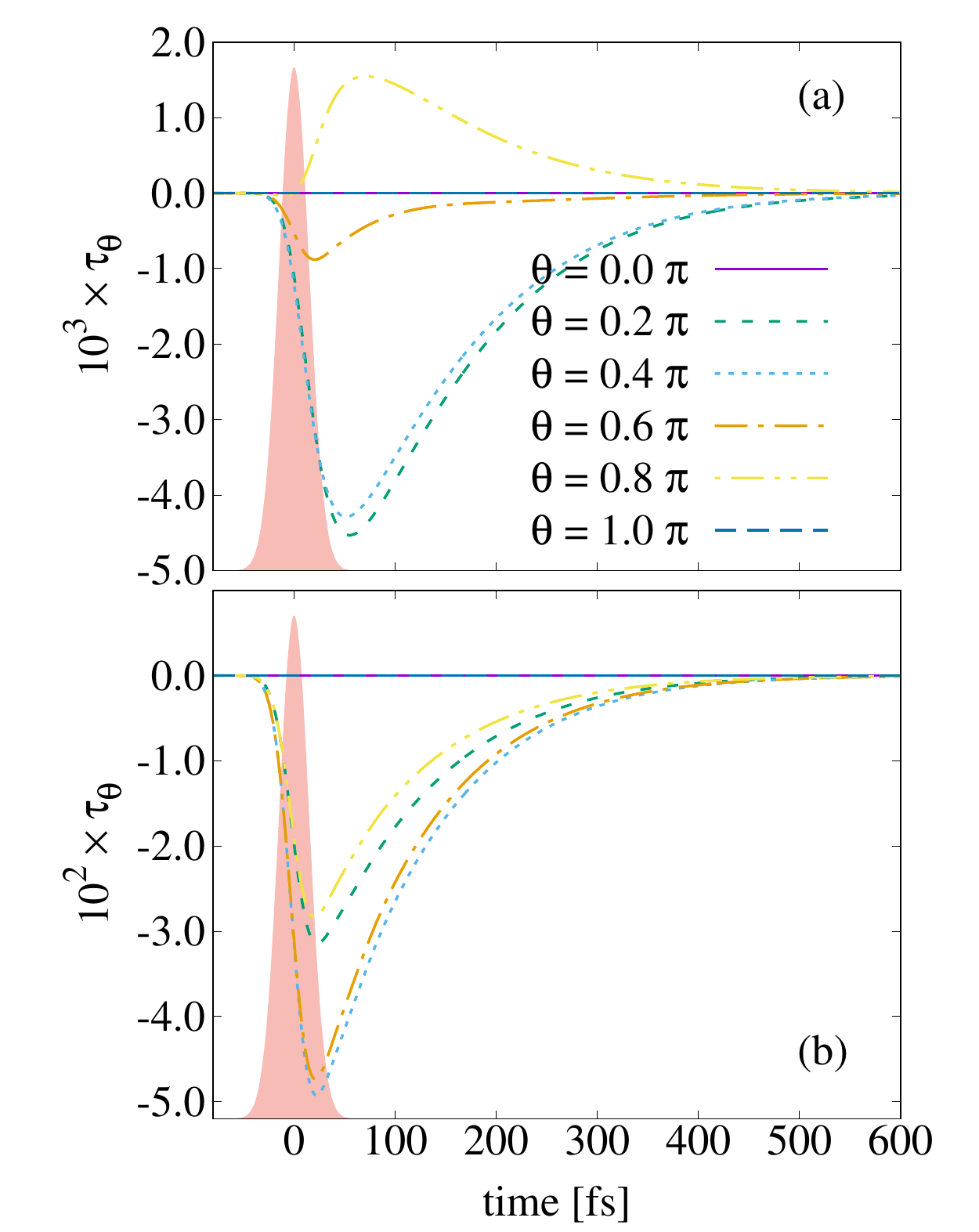}
\caption{\label{Fig:torque} Time dependence of the spin transfer torque acting
on (a) the Ni and (b) the Fe magnetization, excited by a femtosecond laser pulse
per area of $a^2$ with $a$ being the crystal lattice constant.
The STT is calculated for different magnetic configurations given by the angle
between the magnetizations, $\theta$. 
The STT is given in the units of $(\hbar / 2)\, {\rm fs}^{-1}$.
The red filled area corresponds to the time dependence of the laser pulse.} 
\end{figure}

Fig.~\ref{Fig:torque} depicts the time dependence of the spin-transfer
torque action in the magnetic layers at different magnetic configurations
given by the mutual angle, $\theta$. 
As expected, there is no spin-transfer torque in the collinear magnetic configurations. Let us start
with the analysis of the STT action on the Fe magnetizations, shown in Fig.~\ref{Fig:torque}(b).
Both its onset and maximum
correlate closely with the same features of the laser pulse itself, with the delay 
being practically nonexistent for the onset and only about $20\,{\rm fs}$ for the maximum.
This suggests that the torque acting in the Fe layer is primarily generated by the hot electrons, 
which were excited directly by the laser pulse in the Ni layer.
Afterwards it starts to decrease, which is a consequence of hot electrons' relaxation
and decay of the spin currents. The spin-transfer torque in the Fe
layer disappears after $\sim 300\,{\rm fs}$. Moreover, we can see
that the torque in the Fe layer appears to be roughly symmetric with respect to $\theta=\pi/2$.

Let us focus now on the action of the torque on the moment of the Ni layer, shown
in Fig.~\ref{Fig:torque}(a). The STT acting in this layer is mostly
generated by the left-moving electrons coming from the direction of the Fe layer.
These are the hot electrons reflected from the interfaces or the secondary ones excited
in the Fe layer. Therefore, the STT magnitude in
the Ni layer is about $10$ times smaller than the one in Fe. One
can also notice a later onset and longer time required for reaching
the maximum value of the STT than in the Fe layer. 
The origin of the electrons that generate a torque in Ni is recognized
in its angular dependence, too. In contrast to the STT in Fe, the torque in the Ni layer shows a significant
asymmetry with respect to $\theta=\pi/2$. More importantly, when the angle
$\theta$ becomes closer to the AP magnetic configuration, the STT changes
its direction (sign) after some time, which can be seen for curves
calculated for $\theta = 0.8\pi$. 

\begin{figure}[tbp!]
\includegraphics[width=0.9\columnwidth]{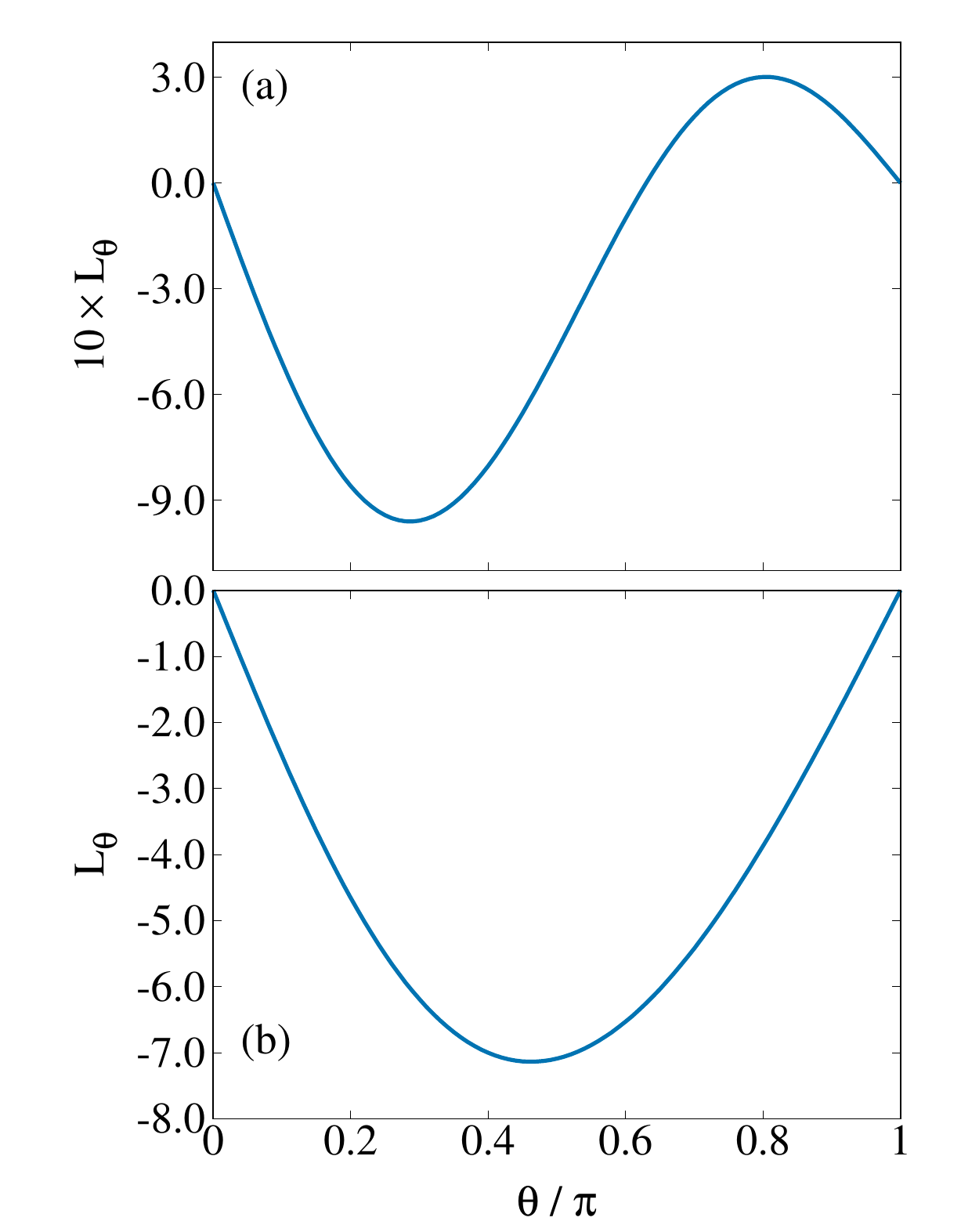} 
\caption{\label{Fig:momentum} Angular dependence of the total angular momentum
due to hot electron-induced STT absorbed in (a) the Ni, and (b) the Fe
magnetic layer per area of $a^2$ with $a$ being the crystal lattice constant. 
$L_\theta$ is given in the units of $\hbar / 2$.}
\end{figure}

As we can see from Fig.~\ref{Fig:torque}, the variation of the STT due to laser-induced hot electron transport is by far faster than the
expected magnetization response, which is usually in the GHz regime.
Therefore, we argue that the total layer magnetization will respond to the
total momentum absorbed in the layer rather than the time dependent STT. 
Therefore, in Fig.~\ref{Fig:momentum} we show the
angular dependence of the total momentum absorbed in the magnetic layer defined as 
\begin{equation}
L_{\theta}(\theta)=\int_{-\infty}^{\infty}{\rm d}t\;\tau_{\theta}(\theta,t)\,.\label{Eq:momentum}
\end{equation}
As expected, $L_{\theta}$ in the Fe layer is a sine-like
function with its maximum close to $\theta = \pi /2$. On the other hand,
$L_{\theta}$ in the Ni layer, which is about $10$ times smaller
than the one in the Fe layer, shows a strongly asymmetric behavior
and changes its sign in certain noncollinear magnetic configuration.

\section{Discussion and conclusions}
\label{Sec:Conclusions}

We presented results of our analysis of laser-induced transport in
magnetic multilayers consisting of Al/Ni/Ru/Fe/Ru layers. In our calculations
we focused on the role of interfaces and noncollinear magnetic configurations.
Using an {\em ab initio} WFM
method~\citep{Xia:prb06,Zwierzycki2008:PSSb}
we have calculated spin- and energy-dependent averaged interface reflections and
transmissions for single interfaces and also for a central part of the multilayer, i.e.\ Ni/Ru/Fe.
We demonstrated that the results of the fully quantum mechanical, coherent calculations 
can be successfully approximated using the semiclassical formula 
for adding transmissions through Ni/Ru and Ru/Fe interfaces (also taken from \emph{ab initio} calculations), 
provided that the multiple reflections inside the spacer are included. Omission of the internal 
reflections leads to a serious underestimation of the total spin-dependent transmissions. 

In order to treat noncollinear magnetic configurations, we have extended
the standard model of superdiffusive spin-dependent transport \citep{Battiato2014_JAP}.
We approximated the central, thin nonmagnetic Ru spacer, separating
two magnetic layers, Ni and Fe, as a single, effective interface with its transport
properties parameterized by the reflection and transmission
probabilities. The averaged scattering probabilities were calculated in a way allowing 
for including both the impact of multiple (non-coherent) scattering at internal interfaces and
the mixing of the spin-channels present for the noncollinear configurations.

The accuracy of our treatment of the central nonmagnetic layer has been
tested in collinear magnetic configurations by comparing the demagnetization
in the magnetic layers calculated in the current approximation (with
zero spacer thickness) to the one calculated using a model with finite
spacer thickness~\citep{Battiato2014_JAP}. There is a qualitative agreement between both models,
however, quantitatively we observe differences. Our approach provides
a slightly smaller demagnetization of the Ni layer. On the other hand, one observes a
stronger effect in the Fe magnetic layer. These effects can be explained by the omission 
of the bulk part of the Ru layer in our model.
More specifically, in the nonzero-thickness-spacer model, the effect in the Fe
layer is reduced due to electron scattering in the Ru spacer during
the superdiffusive transport. This effect is not included in the transport
parameters in the current model with zero-thickness Ru spacer.

Our calculations reproduced the main features of ultrafast demagnetization observed in experiment~\citep{Rudolf2012}, i.e.\ ultrafast demagnetization of Ni and Fe layer in the parallel magnetic
configuration, and ultrafast increase of the magnetization in the Fe
layer in the antiparallel one. In addition to this,
we have shown that there is a smooth transition between negative and positive
change of magnetization in the Fe layer as the angle between 
Ni and Fe magnetization vectors varies from $0$ to $\pi$. 
On the other hand, for the studied multilayer, the demagnetization of the Ni
magnetic layer does not seem to be meaningfully affected by the magnetic configuration. 
Consequently, for a certain noncollinear magnetic configuration, 
one can observe demagnetization in the Ni magnetic layer, 
but effectively no total change of magnetization in the Fe layer.

We also studied the spin-transfer torque acting in both magnetic layers. 
When hot electrons were excited just in Ni, 
the spin current flows mainly from Ni towards the Fe layer and leads to a spin-transfer
torque acting on the Fe magnetization. Our simulations have shown that the
total transverse angular momentum (time-integrated spin-transfer torque)
passing the nonmagnetic spacer depends on the angle between magnetizations as a sine-like function, 
which resembles the regular (``cold") spin-transfer torque in magnetic
spin valves~\citep{Slonczewski1996,Barnas2005}. 
We also observed a spin current flowing from Fe towards the Ni layer. This
spin current consists of electrons reflected from Ru$/$Fe interface and also
of hot electrons excited secondarily in the Fe layer.
Interestingly, the STT changes sign when the angle between the magnetization
of Fe and Ni layers exceeds $\theta \approx 0.6\pi$. The reason for this behavior is that
for these values of $\theta$ the reflected electrons become polarized mostly 
in the direction of the Fe magnetization.
Consequently, the angular dependence of the total transverse angular momentum
passing from Fe to the Ni layer becomes zero at certain noncollinear configurations
and then changes its sign for increased angles.

In summary, the model presented in this paper fills the gap in the current
research of the laser-induced spin transport in multilayered nanostructures. 
On the one hand, we provided a systematical study of the effect 
of a thin nonmagnetic spacer, including its interfaces,
on the laser-pulse-induced demagnetization in the magnetic layers. 
The proper treatment of the interfaces might have a crucial role in 
explanation of the contrasting results of some recent experimental studies~\cite{Rudolf2012,Eschenlohr2017}.
On the other hand, our study goes beyond the standard examination of laser-induced transport
in the collinear magnetic configurations, and allows to study nonequilibrium electronic spin transport in  noncollinear magnetic heterostructures,
which might open new ways toward laser-driven magnetization switching in spintronic devices.

\section*{Acknowledgment}

This work was supported by the European Regional Development Fund
in the IT4Innovations national supercomputing center - path to exascale
project (project number CZ.02.1.01/0.0/0.0/16\_013/0001791) within
the Operational Programme Research, Development and Education, by
the Czech Science Foundation (grant number 19-13659S), by the Swedish
Research Council (VR), and by The Ministry of Education, Youth and
Sports from the Large Infrastructures for Research, Experimental Development
and Innovations project ``e-Infrastructure CZ -- LM2018140''.
This work has furthermore been funded by the European Union’s Horizon2020 Research and Innovation Programme under FET-OPEN Grant Agreement No.\ 863155 (s-Nebula). We acknowledge computational resources provided by the Swedish National Infrastructure for Computing (SNIC), partially funded by the Swedish Research Council through Grant Agreement No.\ 2018-05973.

\bibliographystyle{apsrev4-2}
\bibliography{bibliography}

\begin{thebibliography}{59}%
\makeatletter
\providecommand \@ifxundefined [1]{%
 \@ifx{#1\undefined}
}%
\providecommand \@ifnum [1]{%
 \ifnum #1\expandafter \@firstoftwo
 \else \expandafter \@secondoftwo
 \fi
}%
\providecommand \@ifx [1]{%
 \ifx #1\expandafter \@firstoftwo
 \else \expandafter \@secondoftwo
 \fi
}%
\providecommand \natexlab [1]{#1}%
\providecommand \enquote  [1]{``#1''}%
\providecommand \bibnamefont  [1]{#1}%
\providecommand \bibfnamefont [1]{#1}%
\providecommand \citenamefont [1]{#1}%
\providecommand \href@noop [0]{\@secondoftwo}%
\providecommand \href [0]{\begingroup \@sanitize@url \@href}%
\providecommand \@href[1]{\@@startlink{#1}\@@href}%
\providecommand \@@href[1]{\endgroup#1\@@endlink}%
\providecommand \@sanitize@url [0]{\catcode `\\12\catcode `\$12\catcode
  `\&12\catcode `\#12\catcode `\^12\catcode `\_12\catcode `\%12\relax}%
\providecommand \@@startlink[1]{}%
\providecommand \@@endlink[0]{}%
\providecommand \url  [0]{\begingroup\@sanitize@url \@url }%
\providecommand \@url [1]{\endgroup\@href {#1}{\urlprefix }}%
\providecommand \urlprefix  [0]{URL }%
\providecommand \Eprint [0]{\href }%
\providecommand \doibase [0]{https://doi.org/}%
\providecommand \selectlanguage [0]{\@gobble}%
\providecommand \bibinfo  [0]{\@secondoftwo}%
\providecommand \bibfield  [0]{\@secondoftwo}%
\providecommand \translation [1]{[#1]}%
\providecommand \BibitemOpen [0]{}%
\providecommand \bibitemStop [0]{}%
\providecommand \bibitemNoStop [0]{.\EOS\space}%
\providecommand \EOS [0]{\spacefactor3000\relax}%
\providecommand \BibitemShut  [1]{\csname bibitem#1\endcsname}%
\let\auto@bib@innerbib\@empty
\bibitem [{\citenamefont {Slonczewski}(1996)}]{Slonczewski1996}%
  \BibitemOpen
  \bibfield  {author} {\bibinfo {author} {\bibfnamefont {J.}~\bibnamefont
  {Slonczewski}},\ }\href
  {https://doi.org/https://doi.org/10.1016/0304-8853(96)00062-5} {\bibfield
  {journal} {\bibinfo  {journal} {J. Magn. Magn. Mater.}\ }\textbf {\bibinfo
  {volume} {159}},\ \bibinfo {pages} {L1} (\bibinfo {year} {1996})}\BibitemShut
  {NoStop}%
\bibitem [{\citenamefont {Berger}(1996)}]{Berger1996}%
  \BibitemOpen
  \bibfield  {author} {\bibinfo {author} {\bibfnamefont {L.}~\bibnamefont
  {Berger}},\ }\href {https://doi.org/10.1103/PhysRevB.54.9353} {\bibfield
  {journal} {\bibinfo  {journal} {Phys. Rev. B}\ }\textbf {\bibinfo {volume}
  {54}},\ \bibinfo {pages} {9353} (\bibinfo {year} {1996})}\BibitemShut
  {NoStop}%
\bibitem [{\citenamefont {Miron}\ \emph {et~al.}(2011)\citenamefont {Miron},
  \citenamefont {Garello}, \citenamefont {Gaudin}, \citenamefont {Zermatten},
  \citenamefont {Costache}, \citenamefont {Auffret}, \citenamefont {Bandiera},
  \citenamefont {Rodmacq}, \citenamefont {Schuhl},\ and\ \citenamefont
  {Gambardella}}]{Miron:Nature2011}%
  \BibitemOpen
  \bibfield  {author} {\bibinfo {author} {\bibfnamefont {I.~M.}\ \bibnamefont
  {Miron}}, \bibinfo {author} {\bibfnamefont {K.}~\bibnamefont {Garello}},
  \bibinfo {author} {\bibfnamefont {G.}~\bibnamefont {Gaudin}}, \bibinfo
  {author} {\bibfnamefont {P.-J.}\ \bibnamefont {Zermatten}}, \bibinfo {author}
  {\bibfnamefont {M.~V.}\ \bibnamefont {Costache}}, \bibinfo {author}
  {\bibfnamefont {S.}~\bibnamefont {Auffret}}, \bibinfo {author} {\bibfnamefont
  {S.}~\bibnamefont {Bandiera}}, \bibinfo {author} {\bibfnamefont
  {B.}~\bibnamefont {Rodmacq}}, \bibinfo {author} {\bibfnamefont
  {A.}~\bibnamefont {Schuhl}},\ and\ \bibinfo {author} {\bibfnamefont
  {P.}~\bibnamefont {Gambardella}},\ }\href
  {https://doi.org/10.1038/nature10309} {\bibfield  {journal} {\bibinfo
  {journal} {Nature}\ }\textbf {\bibinfo {volume} {476}},\ \bibinfo {pages}
  {189} (\bibinfo {year} {2011})}\BibitemShut {NoStop}%
\bibitem [{\citenamefont {Liu}\ \emph {et~al.}(2012)\citenamefont {Liu},
  \citenamefont {Pai}, \citenamefont {Li}, \citenamefont {Tseng}, \citenamefont
  {Ralph},\ and\ \citenamefont {Buhrman}}]{Liu:Science2012}%
  \BibitemOpen
  \bibfield  {author} {\bibinfo {author} {\bibfnamefont {L.}~\bibnamefont
  {Liu}}, \bibinfo {author} {\bibfnamefont {C.-F.}\ \bibnamefont {Pai}},
  \bibinfo {author} {\bibfnamefont {Y.}~\bibnamefont {Li}}, \bibinfo {author}
  {\bibfnamefont {H.~W.}\ \bibnamefont {Tseng}}, \bibinfo {author}
  {\bibfnamefont {D.~C.}\ \bibnamefont {Ralph}},\ and\ \bibinfo {author}
  {\bibfnamefont {R.~A.}\ \bibnamefont {Buhrman}},\ }\href
  {https://doi.org/10.1126/science.1218197} {\bibfield  {journal} {\bibinfo
  {journal} {Science}\ }\textbf {\bibinfo {volume} {336}},\ \bibinfo {pages}
  {555} (\bibinfo {year} {2012})}\BibitemShut {NoStop}%
\bibitem [{\citenamefont {Jhuria}\ \emph {et~al.}(2020)\citenamefont {Jhuria},
  \citenamefont {Hohlfeld}, \citenamefont {Pattabi}, \citenamefont {Martin},
  \citenamefont {C{\'{o}}rdova}, \citenamefont {Shi}, \citenamefont {Conte},
  \citenamefont {Petit-Watelot}, \citenamefont {Rojas-Sanchez}, \citenamefont
  {Malinowski}, \citenamefont {Mangin}, \citenamefont {Lema{\^{\i}}tre},
  \citenamefont {Hehn}, \citenamefont {Bokor}, \citenamefont {Wilson},\ and\
  \citenamefont {Gorchon}}]{Jhuria2020_SOTsw_ps}%
  \BibitemOpen
  \bibfield  {author} {\bibinfo {author} {\bibfnamefont {K.}~\bibnamefont
  {Jhuria}}, \bibinfo {author} {\bibfnamefont {J.}~\bibnamefont {Hohlfeld}},
  \bibinfo {author} {\bibfnamefont {A.}~\bibnamefont {Pattabi}}, \bibinfo
  {author} {\bibfnamefont {E.}~\bibnamefont {Martin}}, \bibinfo {author}
  {\bibfnamefont {A.~Y.~A.}\ \bibnamefont {C{\'{o}}rdova}}, \bibinfo {author}
  {\bibfnamefont {X.}~\bibnamefont {Shi}}, \bibinfo {author} {\bibfnamefont
  {R.~L.}\ \bibnamefont {Conte}}, \bibinfo {author} {\bibfnamefont
  {S.}~\bibnamefont {Petit-Watelot}}, \bibinfo {author} {\bibfnamefont {J.~C.}\
  \bibnamefont {Rojas-Sanchez}}, \bibinfo {author} {\bibfnamefont
  {G.}~\bibnamefont {Malinowski}}, \bibinfo {author} {\bibfnamefont
  {S.}~\bibnamefont {Mangin}}, \bibinfo {author} {\bibfnamefont
  {A.}~\bibnamefont {Lema{\^{\i}}tre}}, \bibinfo {author} {\bibfnamefont
  {M.}~\bibnamefont {Hehn}}, \bibinfo {author} {\bibfnamefont {J.}~\bibnamefont
  {Bokor}}, \bibinfo {author} {\bibfnamefont {R.~B.}\ \bibnamefont {Wilson}},\
  and\ \bibinfo {author} {\bibfnamefont {J.}~\bibnamefont {Gorchon}},\ }\href
  {https://doi.org/10.1038/s41928-020-00488-3} {\bibfield  {journal} {\bibinfo
  {journal} {Nature Electronics}\ }\textbf {\bibinfo {volume} {3}},\ \bibinfo
  {pages} {680} (\bibinfo {year} {2020})}\BibitemShut {NoStop}%
\bibitem [{\citenamefont {Hirsch}(1999)}]{Hirsch:PRL1999}%
  \BibitemOpen
  \bibfield  {author} {\bibinfo {author} {\bibfnamefont {J.~E.}\ \bibnamefont
  {Hirsch}},\ }\href {https://doi.org/10.1103/PhysRevLett.83.1834} {\bibfield
  {journal} {\bibinfo  {journal} {Phys. Rev. Lett.}\ }\textbf {\bibinfo
  {volume} {83}},\ \bibinfo {pages} {1834} (\bibinfo {year}
  {1999})}\BibitemShut {NoStop}%
\bibitem [{\citenamefont {Sinova}\ \emph {et~al.}(2015)\citenamefont {Sinova},
  \citenamefont {Valenzuela}, \citenamefont {Wunderlich}, \citenamefont
  {Back},\ and\ \citenamefont {Jungwirth}}]{Sinova:RevModPhys2015}%
  \BibitemOpen
  \bibfield  {author} {\bibinfo {author} {\bibfnamefont {J.}~\bibnamefont
  {Sinova}}, \bibinfo {author} {\bibfnamefont {S.~O.}\ \bibnamefont
  {Valenzuela}}, \bibinfo {author} {\bibfnamefont {J.}~\bibnamefont
  {Wunderlich}}, \bibinfo {author} {\bibfnamefont {C.~H.}\ \bibnamefont
  {Back}},\ and\ \bibinfo {author} {\bibfnamefont {T.}~\bibnamefont
  {Jungwirth}},\ }\href {https://doi.org/10.1103/RevModPhys.87.1213} {\bibfield
   {journal} {\bibinfo  {journal} {Rev. Mod. Phys.}\ }\textbf {\bibinfo
  {volume} {87}},\ \bibinfo {pages} {1213} (\bibinfo {year}
  {2015})}\BibitemShut {NoStop}%
\bibitem [{\citenamefont {Beaurepaire}\ \emph {et~al.}(1996)\citenamefont
  {Beaurepaire}, \citenamefont {Merle}, \citenamefont {Daunois},\ and\
  \citenamefont {Bigot}}]{Beaurepaire1996}%
  \BibitemOpen
  \bibfield  {author} {\bibinfo {author} {\bibfnamefont {E.}~\bibnamefont
  {Beaurepaire}}, \bibinfo {author} {\bibfnamefont {J.-C.}\ \bibnamefont
  {Merle}}, \bibinfo {author} {\bibfnamefont {A.}~\bibnamefont {Daunois}},\
  and\ \bibinfo {author} {\bibfnamefont {J.-Y.}\ \bibnamefont {Bigot}},\ }\href
  {https://doi.org/10.1103/PhysRevLett.76.4250} {\bibfield  {journal} {\bibinfo
   {journal} {Phys. Rev. Lett.}\ }\textbf {\bibinfo {volume} {76}},\ \bibinfo
  {pages} {4250} (\bibinfo {year} {1996})}\BibitemShut {NoStop}%
\bibitem [{\citenamefont {Carva}\ \emph {et~al.}(2014)\citenamefont {Carva},
  \citenamefont {Baláž},\ and\ \citenamefont {Radu}}]{Carva2017_Handbook}%
  \BibitemOpen
  \bibfield  {author} {\bibinfo {author} {\bibfnamefont {K.}~\bibnamefont
  {Carva}}, \bibinfo {author} {\bibfnamefont {P.}~\bibnamefont {Baláž}},\
  and\ \bibinfo {author} {\bibfnamefont {I.}~\bibnamefont {Radu}},\ }in\ \href
  {https://doi.org/10.1016/bs.hmm.2017.09.003} {\emph {\bibinfo {booktitle}
  {Handbook of Magnetic Materials}}},\ Vol.~\bibinfo {volume} {26},\ \bibinfo
  {editor} {edited by\ \bibinfo {editor} {\bibfnamefont {E.}~\bibnamefont
  {Brück}}}\ (\bibinfo  {publisher} {Elsevier},\ \bibinfo {year} {2014})\ pp.\
  \bibinfo {pages} {291--463}\BibitemShut {NoStop}%
\bibitem [{\citenamefont {N{\v{e}}mec}\ \emph {et~al.}(2012)\citenamefont
  {N{\v{e}}mec}, \citenamefont {Rozkotov{\'{a}}}, \citenamefont
  {Tesarov{\'{a}}}, \citenamefont {Troj{\'{a}}nek}, \citenamefont {{De
  Ranieri}}, \citenamefont {Olejn{\'{i}}k}, \citenamefont {Zemen},
  \citenamefont {Nov{\'{a}}k}, \citenamefont {Cukr}, \citenamefont
  {Mal{\'{y}}},\ and\ \citenamefont {Jungwirth}}]{r_12_Nemec_opticalSTT}%
  \BibitemOpen
  \bibfield  {author} {\bibinfo {author} {\bibfnamefont {P.}~\bibnamefont
  {N{\v{e}}mec}}, \bibinfo {author} {\bibfnamefont {E.}~\bibnamefont
  {Rozkotov{\'{a}}}}, \bibinfo {author} {\bibfnamefont {N.}~\bibnamefont
  {Tesarov{\'{a}}}}, \bibinfo {author} {\bibfnamefont {F.}~\bibnamefont
  {Troj{\'{a}}nek}}, \bibinfo {author} {\bibfnamefont {E.}~\bibnamefont {{De
  Ranieri}}}, \bibinfo {author} {\bibfnamefont {K.}~\bibnamefont
  {Olejn{\'{i}}k}}, \bibinfo {author} {\bibfnamefont {J.}~\bibnamefont
  {Zemen}}, \bibinfo {author} {\bibfnamefont {V.}~\bibnamefont {Nov{\'{a}}k}},
  \bibinfo {author} {\bibfnamefont {M.}~\bibnamefont {Cukr}}, \bibinfo {author}
  {\bibfnamefont {P.}~\bibnamefont {Mal{\'{y}}}},\ and\ \bibinfo {author}
  {\bibfnamefont {T.}~\bibnamefont {Jungwirth}},\ }\href
  {https://doi.org/10.1038/nphys2279} {\bibfield  {journal} {\bibinfo
  {journal} {Nat. Phys.}\ }\textbf {\bibinfo {volume} {8}},\ \bibinfo {pages}
  {411} (\bibinfo {year} {2012})}\BibitemShut {NoStop}%
\bibitem [{\citenamefont {Tesarov{\'{a}}}\ \emph {et~al.}(2013)\citenamefont
  {Tesarov{\'{a}}}, \citenamefont {N{\v{e}}mec}, \citenamefont
  {Rozkotov{\'{a}}}, \citenamefont {Zemen}, \citenamefont {Janda},
  \citenamefont {Butkovi{\v{c}}ov{\'{a}}}, \citenamefont {Troj{\'{a}}nek},
  \citenamefont {Olejn{\'{i}}k}, \citenamefont {Nov{\'{a}}k}, \citenamefont
  {Mal{\'{y}}},\ and\ \citenamefont
  {Jungwirth}}]{r_13_Tesarova_Nemec_Observ_SOT}%
  \BibitemOpen
  \bibfield  {author} {\bibinfo {author} {\bibfnamefont {N.}~\bibnamefont
  {Tesarov{\'{a}}}}, \bibinfo {author} {\bibfnamefont {P.}~\bibnamefont
  {N{\v{e}}mec}}, \bibinfo {author} {\bibfnamefont {E.}~\bibnamefont
  {Rozkotov{\'{a}}}}, \bibinfo {author} {\bibfnamefont {J.}~\bibnamefont
  {Zemen}}, \bibinfo {author} {\bibfnamefont {T.}~\bibnamefont {Janda}},
  \bibinfo {author} {\bibfnamefont {D.}~\bibnamefont
  {Butkovi{\v{c}}ov{\'{a}}}}, \bibinfo {author} {\bibfnamefont
  {F.}~\bibnamefont {Troj{\'{a}}nek}}, \bibinfo {author} {\bibfnamefont
  {K.}~\bibnamefont {Olejn{\'{i}}k}}, \bibinfo {author} {\bibfnamefont
  {V.}~\bibnamefont {Nov{\'{a}}k}}, \bibinfo {author} {\bibfnamefont
  {P.}~\bibnamefont {Mal{\'{y}}}},\ and\ \bibinfo {author} {\bibfnamefont
  {T.}~\bibnamefont {Jungwirth}},\ }\href
  {https://doi.org/10.1038/nphoton.2013.76} {\bibfield  {journal} {\bibinfo
  {journal} {Nat. Photon.}\ }\textbf {\bibinfo {volume} {7}},\ \bibinfo {pages}
  {492} (\bibinfo {year} {2013})}\BibitemShut {NoStop}%
\bibitem [{\citenamefont {Malinowski}\ \emph {et~al.}(2008)\citenamefont
  {Malinowski}, \citenamefont {Dalla~Longa}, \citenamefont {Rietjens},
  \citenamefont {Paluskar}, \citenamefont {Huijink}, \citenamefont {Swagten},\
  and\ \citenamefont {Koopmans}}]{Malinowski2008:NatPhys}%
  \BibitemOpen
  \bibfield  {author} {\bibinfo {author} {\bibfnamefont {G.}~\bibnamefont
  {Malinowski}}, \bibinfo {author} {\bibfnamefont {F.}~\bibnamefont
  {Dalla~Longa}}, \bibinfo {author} {\bibfnamefont {J.~H.~H.}\ \bibnamefont
  {Rietjens}}, \bibinfo {author} {\bibfnamefont {P.~V.}\ \bibnamefont
  {Paluskar}}, \bibinfo {author} {\bibfnamefont {R.}~\bibnamefont {Huijink}},
  \bibinfo {author} {\bibfnamefont {H.~J.~M.}\ \bibnamefont {Swagten}},\ and\
  \bibinfo {author} {\bibfnamefont {B.}~\bibnamefont {Koopmans}},\ }\href
  {https://doi.org/10.1038/nphys1092} {\bibfield  {journal} {\bibinfo
  {journal} {Nature Physics}\ }\textbf {\bibinfo {volume} {4}},\ \bibinfo
  {pages} {855} (\bibinfo {year} {2008})}\BibitemShut {NoStop}%
\bibitem [{\citenamefont {Battiato}\ \emph {et~al.}(2010)\citenamefont
  {Battiato}, \citenamefont {Carva},\ and\ \citenamefont
  {Oppeneer}}]{Battiato2010}%
  \BibitemOpen
  \bibfield  {author} {\bibinfo {author} {\bibfnamefont {M.}~\bibnamefont
  {Battiato}}, \bibinfo {author} {\bibfnamefont {K.}~\bibnamefont {Carva}},\
  and\ \bibinfo {author} {\bibfnamefont {P.~M.}\ \bibnamefont {Oppeneer}},\
  }\href {https://link.aps.org/doi/10.1103/PhysRevLett.105.027203} {\bibfield
  {journal} {\bibinfo  {journal} {Phys. Rev. Lett.}\ }\textbf {\bibinfo
  {volume} {105}},\ \bibinfo {pages} {027203} (\bibinfo {year}
  {2010})}\BibitemShut {NoStop}%
\bibitem [{\citenamefont {Battiato}\ \emph {et~al.}(2012)\citenamefont
  {Battiato}, \citenamefont {Carva},\ and\ \citenamefont
  {Oppeneer}}]{Battiato2012}%
  \BibitemOpen
  \bibfield  {author} {\bibinfo {author} {\bibfnamefont {M.}~\bibnamefont
  {Battiato}}, \bibinfo {author} {\bibfnamefont {K.}~\bibnamefont {Carva}},\
  and\ \bibinfo {author} {\bibfnamefont {P.~M.}\ \bibnamefont {Oppeneer}},\
  }\href {https://link.aps.org/doi/10.1103/PhysRevB.86.024404} {\bibfield
  {journal} {\bibinfo  {journal} {Phys. Rev. B}\ }\textbf {\bibinfo {volume}
  {86}},\ \bibinfo {pages} {024404} (\bibinfo {year} {2012})}\BibitemShut
  {NoStop}%
\bibitem [{\citenamefont {Melnikov}\ \emph {et~al.}(2011)\citenamefont
  {Melnikov}, \citenamefont {Razdolski}, \citenamefont {Wehling}, \citenamefont
  {Papaioannou}, \citenamefont {Roddatis}, \citenamefont {Fumagalli},
  \citenamefont {Aktsipetrov}, \citenamefont {Lichtenstein},\ and\
  \citenamefont {Bovensiepen}}]{Melnikov2011}%
  \BibitemOpen
  \bibfield  {author} {\bibinfo {author} {\bibfnamefont {A.}~\bibnamefont
  {Melnikov}}, \bibinfo {author} {\bibfnamefont {I.}~\bibnamefont {Razdolski}},
  \bibinfo {author} {\bibfnamefont {T.~O.}\ \bibnamefont {Wehling}}, \bibinfo
  {author} {\bibfnamefont {E.~T.}\ \bibnamefont {Papaioannou}}, \bibinfo
  {author} {\bibfnamefont {V.}~\bibnamefont {Roddatis}}, \bibinfo {author}
  {\bibfnamefont {P.}~\bibnamefont {Fumagalli}}, \bibinfo {author}
  {\bibfnamefont {O.}~\bibnamefont {Aktsipetrov}}, \bibinfo {author}
  {\bibfnamefont {A.~I.}\ \bibnamefont {Lichtenstein}},\ and\ \bibinfo {author}
  {\bibfnamefont {U.}~\bibnamefont {Bovensiepen}},\ }\href
  {https://doi.org/10.1103/PhysRevLett.107.076601} {\bibfield  {journal}
  {\bibinfo  {journal} {Phys. Rev. Lett.}\ }\textbf {\bibinfo {volume} {107}},\
  \bibinfo {pages} {076601} (\bibinfo {year} {2011})}\BibitemShut {NoStop}%
\bibitem [{\citenamefont {Turgut}\ \emph {et~al.}(2013)\citenamefont {Turgut},
  \citenamefont {La-o vorakiat}, \citenamefont {Shaw}, \citenamefont
  {Grychtol}, \citenamefont {Nembach}, \citenamefont {Rudolf}, \citenamefont
  {Adam}, \citenamefont {Aeschlimann}, \citenamefont {Schneider}, \citenamefont
  {Silva}, \citenamefont {Murnane}, \citenamefont {Kapteyn},\ and\
  \citenamefont {Mathias}}]{Turgut2013}%
  \BibitemOpen
  \bibfield  {author} {\bibinfo {author} {\bibfnamefont {E.}~\bibnamefont
  {Turgut}}, \bibinfo {author} {\bibfnamefont {C.}~\bibnamefont {La-o
  vorakiat}}, \bibinfo {author} {\bibfnamefont {J.~M.}\ \bibnamefont {Shaw}},
  \bibinfo {author} {\bibfnamefont {P.}~\bibnamefont {Grychtol}}, \bibinfo
  {author} {\bibfnamefont {H.~T.}\ \bibnamefont {Nembach}}, \bibinfo {author}
  {\bibfnamefont {D.}~\bibnamefont {Rudolf}}, \bibinfo {author} {\bibfnamefont
  {R.}~\bibnamefont {Adam}}, \bibinfo {author} {\bibfnamefont {M.}~\bibnamefont
  {Aeschlimann}}, \bibinfo {author} {\bibfnamefont {C.~M.}\ \bibnamefont
  {Schneider}}, \bibinfo {author} {\bibfnamefont {T.~J.}\ \bibnamefont
  {Silva}}, \bibinfo {author} {\bibfnamefont {M.~M.}\ \bibnamefont {Murnane}},
  \bibinfo {author} {\bibfnamefont {H.~C.}\ \bibnamefont {Kapteyn}},\ and\
  \bibinfo {author} {\bibfnamefont {S.}~\bibnamefont {Mathias}},\ }\href
  {https://doi.org/10.1103/PhysRevLett.110.197201} {\bibfield  {journal}
  {\bibinfo  {journal} {Phys. Rev. Lett.}\ }\textbf {\bibinfo {volume} {110}},\
  \bibinfo {pages} {197201} (\bibinfo {year} {2013})}\BibitemShut {NoStop}%
\bibitem [{\citenamefont {Schellekens}\ \emph {et~al.}(2014)\citenamefont
  {Schellekens}, \citenamefont {Kuiper}, \citenamefont {de~Wit},\ and\
  \citenamefont {Koopmans}}]{Schellekens2014}%
  \BibitemOpen
  \bibfield  {author} {\bibinfo {author} {\bibfnamefont {A.~J.}\ \bibnamefont
  {Schellekens}}, \bibinfo {author} {\bibfnamefont {K.~C.}\ \bibnamefont
  {Kuiper}}, \bibinfo {author} {\bibfnamefont {R.~R. J.~C.}\ \bibnamefont
  {de~Wit}},\ and\ \bibinfo {author} {\bibfnamefont {B.}~\bibnamefont
  {Koopmans}},\ }\href {http://dx.doi.org/10.1038/ncomms5333} {\bibfield
  {journal} {\bibinfo  {journal} {Nat. Commun.}\ }\textbf {\bibinfo {volume}
  {5}},\ \bibinfo {pages} {4333} (\bibinfo {year} {2014})}\BibitemShut
  {NoStop}%
\bibitem [{\citenamefont {Choi}\ \emph {et~al.}(2014)\citenamefont {Choi},
  \citenamefont {Min}, \citenamefont {Lee},\ and\ \citenamefont
  {Cahill}}]{Choi2014}%
  \BibitemOpen
  \bibfield  {author} {\bibinfo {author} {\bibfnamefont {G.-M.}\ \bibnamefont
  {Choi}}, \bibinfo {author} {\bibfnamefont {B.-C.}\ \bibnamefont {Min}},
  \bibinfo {author} {\bibfnamefont {K.-J.}\ \bibnamefont {Lee}},\ and\ \bibinfo
  {author} {\bibfnamefont {D.~G.}\ \bibnamefont {Cahill}},\ }\href
  {http://dx.doi.org/10.1038/ncomms5334} {\bibfield  {journal} {\bibinfo
  {journal} {Nat. Commun.}\ }\textbf {\bibinfo {volume} {5}},\ \bibinfo {pages}
  {4334} (\bibinfo {year} {2014})}\BibitemShut {NoStop}%
\bibitem [{\citenamefont {Bergeard}\ \emph {et~al.}(2016)\citenamefont
  {Bergeard}, \citenamefont {Hehn}, \citenamefont {Mangin}, \citenamefont
  {Lengaigne}, \citenamefont {Montaigne}, \citenamefont {Lalieu}, \citenamefont
  {Koopmans},\ and\ \citenamefont {Malinowski}}]{Bergeard2016}%
  \BibitemOpen
  \bibfield  {author} {\bibinfo {author} {\bibfnamefont {N.}~\bibnamefont
  {Bergeard}}, \bibinfo {author} {\bibfnamefont {M.}~\bibnamefont {Hehn}},
  \bibinfo {author} {\bibfnamefont {S.}~\bibnamefont {Mangin}}, \bibinfo
  {author} {\bibfnamefont {G.}~\bibnamefont {Lengaigne}}, \bibinfo {author}
  {\bibfnamefont {F.}~\bibnamefont {Montaigne}}, \bibinfo {author}
  {\bibfnamefont {M.~L.~M.}\ \bibnamefont {Lalieu}}, \bibinfo {author}
  {\bibfnamefont {B.}~\bibnamefont {Koopmans}},\ and\ \bibinfo {author}
  {\bibfnamefont {G.}~\bibnamefont {Malinowski}},\ }\href
  {https://link.aps.org/doi/10.1103/PhysRevLett.117.147203} {\bibfield
  {journal} {\bibinfo  {journal} {Phys. Rev. Lett.}\ }\textbf {\bibinfo
  {volume} {117}},\ \bibinfo {pages} {147203} (\bibinfo {year}
  {2016})}\BibitemShut {NoStop}%
\bibitem [{\citenamefont {Hofherr}\ \emph {et~al.}(2017)\citenamefont
  {Hofherr}, \citenamefont {Maldonado}, \citenamefont {Schmitt}, \citenamefont
  {Berritta}, \citenamefont {Bierbrauer}, \citenamefont {Sadashivaiah},
  \citenamefont {Schellekens}, \citenamefont {Koopmans}, \citenamefont {Steil},
  \citenamefont {Cinchetti}, \citenamefont {Stadtm\"uller}, \citenamefont
  {Oppeneer}, \citenamefont {Mathias},\ and\ \citenamefont
  {Aeschlimann}}]{Hofherr2017}%
  \BibitemOpen
  \bibfield  {author} {\bibinfo {author} {\bibfnamefont {M.}~\bibnamefont
  {Hofherr}}, \bibinfo {author} {\bibfnamefont {P.}~\bibnamefont {Maldonado}},
  \bibinfo {author} {\bibfnamefont {O.}~\bibnamefont {Schmitt}}, \bibinfo
  {author} {\bibfnamefont {M.}~\bibnamefont {Berritta}}, \bibinfo {author}
  {\bibfnamefont {U.}~\bibnamefont {Bierbrauer}}, \bibinfo {author}
  {\bibfnamefont {S.}~\bibnamefont {Sadashivaiah}}, \bibinfo {author}
  {\bibfnamefont {A.~J.}\ \bibnamefont {Schellekens}}, \bibinfo {author}
  {\bibfnamefont {B.}~\bibnamefont {Koopmans}}, \bibinfo {author}
  {\bibfnamefont {D.}~\bibnamefont {Steil}}, \bibinfo {author} {\bibfnamefont
  {M.}~\bibnamefont {Cinchetti}}, \bibinfo {author} {\bibfnamefont
  {B.}~\bibnamefont {Stadtm\"uller}}, \bibinfo {author} {\bibfnamefont {P.~M.}\
  \bibnamefont {Oppeneer}}, \bibinfo {author} {\bibfnamefont {S.}~\bibnamefont
  {Mathias}},\ and\ \bibinfo {author} {\bibfnamefont {M.}~\bibnamefont
  {Aeschlimann}},\ }\href {https://doi.org/10.1103/PhysRevB.96.100403}
  {\bibfield  {journal} {\bibinfo  {journal} {Phys. Rev. B}\ }\textbf {\bibinfo
  {volume} {96}},\ \bibinfo {pages} {100403(R)} (\bibinfo {year}
  {2017})}\BibitemShut {NoStop}%
\bibitem [{\citenamefont {Malinowski}\ \emph {et~al.}(2018)\citenamefont
  {Malinowski}, \citenamefont {Bergeard}, \citenamefont {Hehn},\ and\
  \citenamefont {Mangin}}]{Malinowski2018}%
  \BibitemOpen
  \bibfield  {author} {\bibinfo {author} {\bibfnamefont {G.}~\bibnamefont
  {Malinowski}}, \bibinfo {author} {\bibfnamefont {N.}~\bibnamefont
  {Bergeard}}, \bibinfo {author} {\bibfnamefont {M.}~\bibnamefont {Hehn}},\
  and\ \bibinfo {author} {\bibfnamefont {S.}~\bibnamefont {Mangin}},\ }\href
  {https://doi.org/10.1140/epjb/e2018-80555-5} {\bibfield  {journal} {\bibinfo
  {journal} {Eur. Phys. J. B}\ }\textbf {\bibinfo {volume} {91}},\ \bibinfo
  {pages} {98} (\bibinfo {year} {2018})}\BibitemShut {NoStop}%
\bibitem [{\citenamefont {Kumberg}\ \emph {et~al.}(2020)\citenamefont
  {Kumberg}, \citenamefont {Golias}, \citenamefont {Pontius}, \citenamefont
  {Hosseinifar}, \citenamefont {Frischmuth}, \citenamefont {Gelen},
  \citenamefont {Shinwari}, \citenamefont {Thakur}, \citenamefont
  {Sch\"u\ss{}ler-Langeheine}, \citenamefont {Oppeneer},\ and\ \citenamefont
  {Kuch}}]{Kumberg2020}%
  \BibitemOpen
  \bibfield  {author} {\bibinfo {author} {\bibfnamefont {I.}~\bibnamefont
  {Kumberg}}, \bibinfo {author} {\bibfnamefont {E.}~\bibnamefont {Golias}},
  \bibinfo {author} {\bibfnamefont {N.}~\bibnamefont {Pontius}}, \bibinfo
  {author} {\bibfnamefont {R.}~\bibnamefont {Hosseinifar}}, \bibinfo {author}
  {\bibfnamefont {K.}~\bibnamefont {Frischmuth}}, \bibinfo {author}
  {\bibfnamefont {I.}~\bibnamefont {Gelen}}, \bibinfo {author} {\bibfnamefont
  {T.}~\bibnamefont {Shinwari}}, \bibinfo {author} {\bibfnamefont
  {S.}~\bibnamefont {Thakur}}, \bibinfo {author} {\bibfnamefont
  {C.}~\bibnamefont {Sch\"u\ss{}ler-Langeheine}}, \bibinfo {author}
  {\bibfnamefont {P.~M.}\ \bibnamefont {Oppeneer}},\ and\ \bibinfo {author}
  {\bibfnamefont {W.}~\bibnamefont {Kuch}},\ }\href
  {https://doi.org/10.1103/PhysRevB.102.214418} {\bibfield  {journal} {\bibinfo
   {journal} {Phys. Rev. B}\ }\textbf {\bibinfo {volume} {102}},\ \bibinfo
  {pages} {214418} (\bibinfo {year} {2020})}\BibitemShut {NoStop}%
\bibitem [{\citenamefont {K\"uhne}\ \emph {et~al.}(2022)\citenamefont
  {K\"uhne}, \citenamefont {Beyazit}, \citenamefont {Sothmann}, \citenamefont
  {Jayabalan}, \citenamefont {Diesing}, \citenamefont {Zhou},\ and\
  \citenamefont {Bovensiepen}}]{Kuhne2022}%
  \BibitemOpen
  \bibfield  {author} {\bibinfo {author} {\bibfnamefont {F.}~\bibnamefont
  {K\"uhne}}, \bibinfo {author} {\bibfnamefont {Y.}~\bibnamefont {Beyazit}},
  \bibinfo {author} {\bibfnamefont {B.}~\bibnamefont {Sothmann}}, \bibinfo
  {author} {\bibfnamefont {J.}~\bibnamefont {Jayabalan}}, \bibinfo {author}
  {\bibfnamefont {D.}~\bibnamefont {Diesing}}, \bibinfo {author} {\bibfnamefont
  {P.}~\bibnamefont {Zhou}},\ and\ \bibinfo {author} {\bibfnamefont
  {U.}~\bibnamefont {Bovensiepen}},\ }\href
  {https://doi.org/10.1103/PhysRevResearch.4.033239} {\bibfield  {journal}
  {\bibinfo  {journal} {Phys. Rev. Research}\ }\textbf {\bibinfo {volume}
  {4}},\ \bibinfo {pages} {033239} (\bibinfo {year} {2022})}\BibitemShut
  {NoStop}%
\bibitem [{\citenamefont {Iihama}\ \emph {et~al.}(2018)\citenamefont {Iihama},
  \citenamefont {Xu}, \citenamefont {Deb}, \citenamefont {Malinowski},
  \citenamefont {Hehn}, \citenamefont {Gorchon}, \citenamefont {Fullerton},\
  and\ \citenamefont {Mangin}}]{Iihama2018_SingleShot_AOS_HotEl}%
  \BibitemOpen
  \bibfield  {author} {\bibinfo {author} {\bibfnamefont {S.}~\bibnamefont
  {Iihama}}, \bibinfo {author} {\bibfnamefont {Y.}~\bibnamefont {Xu}}, \bibinfo
  {author} {\bibfnamefont {M.}~\bibnamefont {Deb}}, \bibinfo {author}
  {\bibfnamefont {G.}~\bibnamefont {Malinowski}}, \bibinfo {author}
  {\bibfnamefont {M.}~\bibnamefont {Hehn}}, \bibinfo {author} {\bibfnamefont
  {J.}~\bibnamefont {Gorchon}}, \bibinfo {author} {\bibfnamefont {E.~E.}\
  \bibnamefont {Fullerton}},\ and\ \bibinfo {author} {\bibfnamefont
  {S.}~\bibnamefont {Mangin}},\ }\href {https://doi.org/10.1002/adma.201804004}
  {\bibfield  {journal} {\bibinfo  {journal} {Advanced Materials}\ ,\ \bibinfo
  {pages} {1804004}} (\bibinfo {year} {2018})}\BibitemShut {NoStop}%
\bibitem [{\citenamefont {Remy}\ \emph {et~al.}(2020)\citenamefont {Remy},
  \citenamefont {Igarashi}, \citenamefont {Iihama}, \citenamefont {Malinowski},
  \citenamefont {Hehn}, \citenamefont {Gorchon}, \citenamefont {Hohlfeld},
  \citenamefont {Fukami}, \citenamefont {Ohno},\ and\ \citenamefont
  {Mangin}}]{Remy2020_UltraSC_SingleSwitch}%
  \BibitemOpen
  \bibfield  {author} {\bibinfo {author} {\bibfnamefont {Q.}~\bibnamefont
  {Remy}}, \bibinfo {author} {\bibfnamefont {J.}~\bibnamefont {Igarashi}},
  \bibinfo {author} {\bibfnamefont {S.}~\bibnamefont {Iihama}}, \bibinfo
  {author} {\bibfnamefont {G.}~\bibnamefont {Malinowski}}, \bibinfo {author}
  {\bibfnamefont {M.}~\bibnamefont {Hehn}}, \bibinfo {author} {\bibfnamefont
  {J.}~\bibnamefont {Gorchon}}, \bibinfo {author} {\bibfnamefont
  {J.}~\bibnamefont {Hohlfeld}}, \bibinfo {author} {\bibfnamefont
  {S.}~\bibnamefont {Fukami}}, \bibinfo {author} {\bibfnamefont
  {H.}~\bibnamefont {Ohno}},\ and\ \bibinfo {author} {\bibfnamefont
  {S.}~\bibnamefont {Mangin}},\ }\href {https://doi.org/10.1002/advs.202001996}
  {\bibfield  {journal} {\bibinfo  {journal} {Advanced Science}\ ,\ \bibinfo
  {pages} {2001996}} (\bibinfo {year} {2020})}\BibitemShut {NoStop}%
\bibitem [{\citenamefont {Rudolf}\ \emph {et~al.}(2012)\citenamefont {Rudolf},
  \citenamefont {La-o vorakiat}, \citenamefont {Battiato}, \citenamefont
  {Adam}, \citenamefont {Shaw}, \citenamefont {Turgut}, \citenamefont
  {Maldonado}, \citenamefont {Mathias}, \citenamefont {Grychtol}, \citenamefont
  {Nembach}, \citenamefont {Silva}, \citenamefont {Aeschlimann}, \citenamefont
  {Kapteyn}, \citenamefont {Murnane}, \citenamefont {Schneider},\ and\
  \citenamefont {Oppeneer}}]{Rudolf2012}%
  \BibitemOpen
  \bibfield  {author} {\bibinfo {author} {\bibfnamefont {D.}~\bibnamefont
  {Rudolf}}, \bibinfo {author} {\bibfnamefont {C.}~\bibnamefont {La-o
  vorakiat}}, \bibinfo {author} {\bibfnamefont {M.}~\bibnamefont {Battiato}},
  \bibinfo {author} {\bibfnamefont {R.}~\bibnamefont {Adam}}, \bibinfo {author}
  {\bibfnamefont {J.~M.}\ \bibnamefont {Shaw}}, \bibinfo {author}
  {\bibfnamefont {E.}~\bibnamefont {Turgut}}, \bibinfo {author} {\bibfnamefont
  {P.}~\bibnamefont {Maldonado}}, \bibinfo {author} {\bibfnamefont
  {S.}~\bibnamefont {Mathias}}, \bibinfo {author} {\bibfnamefont
  {P.}~\bibnamefont {Grychtol}}, \bibinfo {author} {\bibfnamefont {H.~T.}\
  \bibnamefont {Nembach}}, \bibinfo {author} {\bibfnamefont {T.~J.}\
  \bibnamefont {Silva}}, \bibinfo {author} {\bibfnamefont {M.}~\bibnamefont
  {Aeschlimann}}, \bibinfo {author} {\bibfnamefont {H.~C.}\ \bibnamefont
  {Kapteyn}}, \bibinfo {author} {\bibfnamefont {M.~M.}\ \bibnamefont
  {Murnane}}, \bibinfo {author} {\bibfnamefont {C.~M.}\ \bibnamefont
  {Schneider}},\ and\ \bibinfo {author} {\bibfnamefont {P.~M.}\ \bibnamefont
  {Oppeneer}},\ }\href {https://doi.org/10.1038/ncomms2029} {\bibfield
  {journal} {\bibinfo  {journal} {Nat. Commun.}\ }\textbf {\bibinfo {volume}
  {3}},\ \bibinfo {pages} {1037} (\bibinfo {year} {2012})}\BibitemShut
  {NoStop}%
\bibitem [{\citenamefont {Eschenlohr}\ \emph {et~al.}(2017)\citenamefont
  {Eschenlohr}, \citenamefont {Persichetti}, \citenamefont {Kachel},
  \citenamefont {Gabureac}, \citenamefont {Gambardella},\ and\ \citenamefont
  {Stamm}}]{Eschenlohr2017}%
  \BibitemOpen
  \bibfield  {author} {\bibinfo {author} {\bibfnamefont {A.}~\bibnamefont
  {Eschenlohr}}, \bibinfo {author} {\bibfnamefont {L.}~\bibnamefont
  {Persichetti}}, \bibinfo {author} {\bibfnamefont {T.}~\bibnamefont {Kachel}},
  \bibinfo {author} {\bibfnamefont {M.}~\bibnamefont {Gabureac}}, \bibinfo
  {author} {\bibfnamefont {P.}~\bibnamefont {Gambardella}},\ and\ \bibinfo
  {author} {\bibfnamefont {C.}~\bibnamefont {Stamm}},\ }\href
  {https://doi.org/10.1088/1361-648X/aa7dd3} {\bibfield  {journal} {\bibinfo
  {journal} {J. Phys.: Condens. Matter}\ }\textbf {\bibinfo {volume} {29}},\
  \bibinfo {pages} {384002} (\bibinfo {year} {2017})}\BibitemShut {NoStop}%
\bibitem [{\citenamefont {Siegrist}\ \emph {et~al.}(2019)\citenamefont
  {Siegrist}, \citenamefont {Gessner}, \citenamefont {Ossiander}, \citenamefont
  {Denker}, \citenamefont {Chang}, \citenamefont {Schröder}, \citenamefont
  {Guggenmos}, \citenamefont {Cui}, \citenamefont {Walowski}, \citenamefont
  {Martens}, \citenamefont {Dewhurst}, \citenamefont {Kleineberg},
  \citenamefont {Münzenberg}, \citenamefont {Sharma},\ and\ \citenamefont
  {Schultze}}]{Siegrist2019_Nat_LightMag_CohControl}%
  \BibitemOpen
  \bibfield  {author} {\bibinfo {author} {\bibfnamefont {F.}~\bibnamefont
  {Siegrist}}, \bibinfo {author} {\bibfnamefont {J.~A.}\ \bibnamefont
  {Gessner}}, \bibinfo {author} {\bibfnamefont {M.}~\bibnamefont {Ossiander}},
  \bibinfo {author} {\bibfnamefont {C.}~\bibnamefont {Denker}}, \bibinfo
  {author} {\bibfnamefont {Y.-P.}\ \bibnamefont {Chang}}, \bibinfo {author}
  {\bibfnamefont {M.~C.}\ \bibnamefont {Schröder}}, \bibinfo {author}
  {\bibfnamefont {A.}~\bibnamefont {Guggenmos}}, \bibinfo {author}
  {\bibfnamefont {Y.}~\bibnamefont {Cui}}, \bibinfo {author} {\bibfnamefont
  {J.}~\bibnamefont {Walowski}}, \bibinfo {author} {\bibfnamefont
  {U.}~\bibnamefont {Martens}}, \bibinfo {author} {\bibfnamefont {J.~K.}\
  \bibnamefont {Dewhurst}}, \bibinfo {author} {\bibfnamefont {U.}~\bibnamefont
  {Kleineberg}}, \bibinfo {author} {\bibfnamefont {M.}~\bibnamefont
  {Münzenberg}}, \bibinfo {author} {\bibfnamefont {S.}~\bibnamefont
  {Sharma}},\ and\ \bibinfo {author} {\bibfnamefont {M.}~\bibnamefont
  {Schultze}},\ }\href {https://doi.org/10.1038/s41586-019-1333-x} {\bibfield
  {journal} {\bibinfo  {journal} {Nature}\ }\textbf {\bibinfo {volume} {571}},\
  \bibinfo {pages} {240} (\bibinfo {year} {2019})}\BibitemShut {NoStop}%
\bibitem [{\citenamefont {Vaskivskyi}\ \emph {et~al.}(2021)\citenamefont
  {Vaskivskyi}, \citenamefont {Malik}, \citenamefont {Salemi}, \citenamefont
  {Turenne}, \citenamefont {Knut}, \citenamefont {Brock}, \citenamefont
  {Stefanuik}, \citenamefont {Söderström}, \citenamefont {Carva},
  \citenamefont {Fullerton}, \citenamefont {Oppeneer}, \citenamefont {Karis},\
  and\ \citenamefont {Dürr}}]{Vaskivskyi2021_CoPt}%
  \BibitemOpen
  \bibfield  {author} {\bibinfo {author} {\bibfnamefont {I.}~\bibnamefont
  {Vaskivskyi}}, \bibinfo {author} {\bibfnamefont {R.~S.}\ \bibnamefont
  {Malik}}, \bibinfo {author} {\bibfnamefont {L.}~\bibnamefont {Salemi}},
  \bibinfo {author} {\bibfnamefont {D.}~\bibnamefont {Turenne}}, \bibinfo
  {author} {\bibfnamefont {R.}~\bibnamefont {Knut}}, \bibinfo {author}
  {\bibfnamefont {J.}~\bibnamefont {Brock}}, \bibinfo {author} {\bibfnamefont
  {R.}~\bibnamefont {Stefanuik}}, \bibinfo {author} {\bibfnamefont
  {J.}~\bibnamefont {Söderström}}, \bibinfo {author} {\bibfnamefont
  {K.}~\bibnamefont {Carva}}, \bibinfo {author} {\bibfnamefont {E.~E.}\
  \bibnamefont {Fullerton}}, \bibinfo {author} {\bibfnamefont {P.~M.}\
  \bibnamefont {Oppeneer}}, \bibinfo {author} {\bibfnamefont {O.}~\bibnamefont
  {Karis}},\ and\ \bibinfo {author} {\bibfnamefont {H.~A.}\ \bibnamefont
  {Dürr}},\ }\href {https://doi.org/10.1021/acs.jpcc.1c02311} {\bibfield
  {journal} {\bibinfo  {journal} {J. Phys. Chem. C}\ }\textbf {\bibinfo
  {volume} {125}},\ \bibinfo {pages} {11714} (\bibinfo {year}
  {2021})}\BibitemShut {NoStop}%
\bibitem [{\citenamefont {Hennes}\ \emph {et~al.}(2022)\citenamefont {Hennes},
  \citenamefont {Lambert}, \citenamefont {Chardonnet}, \citenamefont
  {Delaunay}, \citenamefont {Chiuzb{\u{a}}ian}, \citenamefont {Jal},\ and\
  \citenamefont {Vodungbo}}]{Hennes2022_ElemSelDemag_CoPt}%
  \BibitemOpen
  \bibfield  {author} {\bibinfo {author} {\bibfnamefont {M.}~\bibnamefont
  {Hennes}}, \bibinfo {author} {\bibfnamefont {G.}~\bibnamefont {Lambert}},
  \bibinfo {author} {\bibfnamefont {V.}~\bibnamefont {Chardonnet}}, \bibinfo
  {author} {\bibfnamefont {R.}~\bibnamefont {Delaunay}}, \bibinfo {author}
  {\bibfnamefont {G.~S.}\ \bibnamefont {Chiuzb{\u{a}}ian}}, \bibinfo {author}
  {\bibfnamefont {E.}~\bibnamefont {Jal}},\ and\ \bibinfo {author}
  {\bibfnamefont {B.}~\bibnamefont {Vodungbo}},\ }\href
  {https://doi.org/10.1063/5.0080275} {\bibfield  {journal} {\bibinfo
  {journal} {Applied Physics Letters}\ }\textbf {\bibinfo {volume} {120}},\
  \bibinfo {pages} {072408} (\bibinfo {year} {2022})}\BibitemShut {NoStop}%
\bibitem [{\citenamefont {Carva}(2014)}]{Carva2014_NPhys}%
  \BibitemOpen
  \bibfield  {author} {\bibinfo {author} {\bibfnamefont {K.}~\bibnamefont
  {Carva}},\ }\href {http://dx.doi.org/10.1038/nphys3057} {\bibfield  {journal}
  {\bibinfo  {journal} {Nat. Phys.}\ }\textbf {\bibinfo {volume} {10}},\
  \bibinfo {pages} {552} (\bibinfo {year} {2014})}\BibitemShut {NoStop}%
\bibitem [{\citenamefont {Razdolski}\ \emph {et~al.}(2017)\citenamefont
  {Razdolski}, \citenamefont {Alekhin}, \citenamefont {Ilin}, \citenamefont
  {Meyburg}, \citenamefont {Roddatis}, \citenamefont {Diesing}, \citenamefont
  {Bovensiepen},\ and\ \citenamefont {Melnikov}}]{Razdolski2017}%
  \BibitemOpen
  \bibfield  {author} {\bibinfo {author} {\bibfnamefont {I.}~\bibnamefont
  {Razdolski}}, \bibinfo {author} {\bibfnamefont {A.}~\bibnamefont {Alekhin}},
  \bibinfo {author} {\bibfnamefont {N.}~\bibnamefont {Ilin}}, \bibinfo {author}
  {\bibfnamefont {J.~P.}\ \bibnamefont {Meyburg}}, \bibinfo {author}
  {\bibfnamefont {V.}~\bibnamefont {Roddatis}}, \bibinfo {author}
  {\bibfnamefont {D.}~\bibnamefont {Diesing}}, \bibinfo {author} {\bibfnamefont
  {U.}~\bibnamefont {Bovensiepen}},\ and\ \bibinfo {author} {\bibfnamefont
  {A.}~\bibnamefont {Melnikov}},\ }\href {https://doi.org/10.1038/ncomms15007}
  {\bibfield  {journal} {\bibinfo  {journal} {Nat. Commun.}\ }\textbf {\bibinfo
  {volume} {8}},\ \bibinfo {pages} {15007} (\bibinfo {year}
  {2017})}\BibitemShut {NoStop}%
\bibitem [{\citenamefont {Lalieu}\ \emph {et~al.}(2019)\citenamefont {Lalieu},
  \citenamefont {Lavrijsen}, \citenamefont {Duine},\ and\ \citenamefont
  {Koopmans}}]{Lalieu2019_THz_SpW_Noncol}%
  \BibitemOpen
  \bibfield  {author} {\bibinfo {author} {\bibfnamefont {M.~L.~M.}\
  \bibnamefont {Lalieu}}, \bibinfo {author} {\bibfnamefont {R.}~\bibnamefont
  {Lavrijsen}}, \bibinfo {author} {\bibfnamefont {R.~A.}\ \bibnamefont
  {Duine}},\ and\ \bibinfo {author} {\bibfnamefont {B.}~\bibnamefont
  {Koopmans}},\ }\href {https://doi.org/10.1103/physrevb.99.184439} {\bibfield
  {journal} {\bibinfo  {journal} {Physical Review B}\ }\textbf {\bibinfo
  {volume} {99}},\ \bibinfo {pages} {184439} (\bibinfo {year}
  {2019})}\BibitemShut {NoStop}%
\bibitem [{\citenamefont {Bal{\'{a}}{\v{z}}}\ \emph {et~al.}(2018)\citenamefont
  {Bal{\'{a}}{\v{z}}}, \citenamefont {{\v{Z}}onda}, \citenamefont {Carva},
  \citenamefont {Maldonado},\ and\ \citenamefont {Oppeneer}}]{Balaz2018:JPCM}%
  \BibitemOpen
  \bibfield  {author} {\bibinfo {author} {\bibfnamefont {P.}~\bibnamefont
  {Bal{\'{a}}{\v{z}}}}, \bibinfo {author} {\bibfnamefont {M.}~\bibnamefont
  {{\v{Z}}onda}}, \bibinfo {author} {\bibfnamefont {K.}~\bibnamefont {Carva}},
  \bibinfo {author} {\bibfnamefont {P.}~\bibnamefont {Maldonado}},\ and\
  \bibinfo {author} {\bibfnamefont {P.~M.}\ \bibnamefont {Oppeneer}},\ }\href
  {http://stacks.iop.org/0953-8984/30/i=11/a=115801?key=crossref.a8eb47a75e2024e478da6bc0f5822790}
  {\bibfield  {journal} {\bibinfo  {journal} {J. Phys.: Condens. Matter}\
  }\textbf {\bibinfo {volume} {30}},\ \bibinfo {pages} {115801} (\bibinfo
  {year} {2018})}\BibitemShut {NoStop}%
\bibitem [{\citenamefont {Lalieu}\ \emph {et~al.}(2017)\citenamefont {Lalieu},
  \citenamefont {Peeters}, \citenamefont {Haenen}, \citenamefont {Lavrijsen},\
  and\ \citenamefont {Koopmans}}]{Lalieu2017}%
  \BibitemOpen
  \bibfield  {author} {\bibinfo {author} {\bibfnamefont {M.~L.~M.}\
  \bibnamefont {Lalieu}}, \bibinfo {author} {\bibfnamefont {M.~J.~G.}\
  \bibnamefont {Peeters}}, \bibinfo {author} {\bibfnamefont {S.~R.~R.}\
  \bibnamefont {Haenen}}, \bibinfo {author} {\bibfnamefont {R.}~\bibnamefont
  {Lavrijsen}},\ and\ \bibinfo {author} {\bibfnamefont {B.}~\bibnamefont
  {Koopmans}},\ }\href {https://link.aps.org/doi/10.1103/PhysRevB.96.220411}
  {\bibfield  {journal} {\bibinfo  {journal} {Phys. Rev. B}\ }\textbf {\bibinfo
  {volume} {96}},\ \bibinfo {pages} {220411(R)} (\bibinfo {year}
  {2017})}\BibitemShut {NoStop}%
\bibitem [{\citenamefont {Ulrichs}\ and\ \citenamefont
  {Razdolski}(2018)}]{Ulrichs2018}%
  \BibitemOpen
  \bibfield  {author} {\bibinfo {author} {\bibfnamefont {H.}~\bibnamefont
  {Ulrichs}}\ and\ \bibinfo {author} {\bibfnamefont {I.}~\bibnamefont
  {Razdolski}},\ }\href {https://doi.org/10.1103/PhysRevB.98.054429} {\bibfield
   {journal} {\bibinfo  {journal} {Phys. Rev. B}\ }\textbf {\bibinfo {volume}
  {98}},\ \bibinfo {pages} {054429} (\bibinfo {year} {2018})}\BibitemShut
  {NoStop}%
\bibitem [{\citenamefont {Ritzmann}\ \emph {et~al.}(2020)\citenamefont
  {Ritzmann}, \citenamefont {Bal\'a\v{z}}, \citenamefont {Maldonado},
  \citenamefont {Carva},\ and\ \citenamefont {Oppeneer}}]{Ritzmann2020:PRB}%
  \BibitemOpen
  \bibfield  {author} {\bibinfo {author} {\bibfnamefont {U.}~\bibnamefont
  {Ritzmann}}, \bibinfo {author} {\bibfnamefont {P.}~\bibnamefont
  {Bal\'a\v{z}}}, \bibinfo {author} {\bibfnamefont {P.}~\bibnamefont
  {Maldonado}}, \bibinfo {author} {\bibfnamefont {K.}~\bibnamefont {Carva}},\
  and\ \bibinfo {author} {\bibfnamefont {P.~M.}\ \bibnamefont {Oppeneer}},\
  }\href {https://doi.org/10.1103/PhysRevB.101.174427} {\bibfield  {journal}
  {\bibinfo  {journal} {Phys. Rev. B}\ }\textbf {\bibinfo {volume} {101}},\
  \bibinfo {pages} {174427} (\bibinfo {year} {2020})}\BibitemShut {NoStop}%
\bibitem [{\citenamefont {Bal\'a\v{z}}\ \emph {et~al.}(2020)\citenamefont
  {Bal\'a\v{z}}, \citenamefont {Carva}, \citenamefont {Ritzmann}, \citenamefont
  {Maldonado},\ and\ \citenamefont {Oppeneer}}]{Balaz2020:PRB}%
  \BibitemOpen
  \bibfield  {author} {\bibinfo {author} {\bibfnamefont {P.}~\bibnamefont
  {Bal\'a\v{z}}}, \bibinfo {author} {\bibfnamefont {K.}~\bibnamefont {Carva}},
  \bibinfo {author} {\bibfnamefont {U.}~\bibnamefont {Ritzmann}}, \bibinfo
  {author} {\bibfnamefont {P.}~\bibnamefont {Maldonado}},\ and\ \bibinfo
  {author} {\bibfnamefont {P.~M.}\ \bibnamefont {Oppeneer}},\ }\href
  {https://doi.org/10.1103/PhysRevB.101.174418} {\bibfield  {journal} {\bibinfo
   {journal} {Phys. Rev. B}\ }\textbf {\bibinfo {volume} {101}},\ \bibinfo
  {pages} {174418} (\bibinfo {year} {2020})}\BibitemShut {NoStop}%
\bibitem [{\citenamefont {Bergeard}\ \emph {et~al.}(2020)\citenamefont
  {Bergeard}, \citenamefont {Hehn}, \citenamefont {Carva}, \citenamefont
  {Bal{\'{a}}{\v{z}}}, \citenamefont {Mangin},\ and\ \citenamefont
  {Malinowski}}]{Bergeard2020_Tailor_HotEl}%
  \BibitemOpen
  \bibfield  {author} {\bibinfo {author} {\bibfnamefont {N.}~\bibnamefont
  {Bergeard}}, \bibinfo {author} {\bibfnamefont {M.}~\bibnamefont {Hehn}},
  \bibinfo {author} {\bibfnamefont {K.}~\bibnamefont {Carva}}, \bibinfo
  {author} {\bibfnamefont {P.}~\bibnamefont {Bal{\'{a}}{\v{z}}}}, \bibinfo
  {author} {\bibfnamefont {S.}~\bibnamefont {Mangin}},\ and\ \bibinfo {author}
  {\bibfnamefont {G.}~\bibnamefont {Malinowski}},\ }\href
  {https://doi.org/10.1063/5.0018502} {\bibfield  {journal} {\bibinfo
  {journal} {Applied Physics Letters}\ }\textbf {\bibinfo {volume} {117}},\
  \bibinfo {pages} {222408} (\bibinfo {year} {2020})}\BibitemShut {NoStop}%
\bibitem [{\citenamefont {Shalaby}\ \emph {et~al.}(2018)\citenamefont
  {Shalaby}, \citenamefont {Donges}, \citenamefont {Carva}, \citenamefont
  {Allenspach}, \citenamefont {Oppeneer}, \citenamefont {Nowak},\ and\
  \citenamefont {Hauri}}]{Shalaby2018_co_CohIncohDyn_THz}%
  \BibitemOpen
  \bibfield  {author} {\bibinfo {author} {\bibfnamefont {M.}~\bibnamefont
  {Shalaby}}, \bibinfo {author} {\bibfnamefont {A.}~\bibnamefont {Donges}},
  \bibinfo {author} {\bibfnamefont {K.}~\bibnamefont {Carva}}, \bibinfo
  {author} {\bibfnamefont {R.}~\bibnamefont {Allenspach}}, \bibinfo {author}
  {\bibfnamefont {P.~M.}\ \bibnamefont {Oppeneer}}, \bibinfo {author}
  {\bibfnamefont {U.}~\bibnamefont {Nowak}},\ and\ \bibinfo {author}
  {\bibfnamefont {C.~P.}\ \bibnamefont {Hauri}},\ }\href
  {https://doi.org/10.1103/physrevb.98.014405} {\bibfield  {journal} {\bibinfo
  {journal} {Phys. Rev. B}\ }\textbf {\bibinfo {volume} {98}},\ \bibinfo
  {pages} {014405} (\bibinfo {year} {2018})}\BibitemShut {NoStop}%
\bibitem [{\citenamefont {Berritta}\ \emph {et~al.}(2016)\citenamefont
  {Berritta}, \citenamefont {Mondal}, \citenamefont {Carva},\ and\
  \citenamefont {Oppeneer}}]{Berritta2016}%
  \BibitemOpen
  \bibfield  {author} {\bibinfo {author} {\bibfnamefont {M.}~\bibnamefont
  {Berritta}}, \bibinfo {author} {\bibfnamefont {R.}~\bibnamefont {Mondal}},
  \bibinfo {author} {\bibfnamefont {K.}~\bibnamefont {Carva}},\ and\ \bibinfo
  {author} {\bibfnamefont {P.~M.}\ \bibnamefont {Oppeneer}},\ }\href
  {https://doi.org/10.1103/PhysRevLett.117.137203} {\bibfield  {journal}
  {\bibinfo  {journal} {Phys. Rev. Lett.}\ }\textbf {\bibinfo {volume} {117}},\
  \bibinfo {pages} {137203} (\bibinfo {year} {2016})}\BibitemShut {NoStop}%
\bibitem [{\citenamefont {Kerber}\ \emph {et~al.}(2020)\citenamefont {Kerber},
  \citenamefont {Ksenzov}, \citenamefont {Freimuth}, \citenamefont {Capotondi},
  \citenamefont {Pedersoli}, \citenamefont {Lopez-Quintas}, \citenamefont
  {Seng}, \citenamefont {Cramer}, \citenamefont {Litzius}, \citenamefont
  {Lacour}, \citenamefont {Zabel}, \citenamefont {Mokrousov}, \citenamefont
  {Kläui},\ and\ \citenamefont {Gutt}}]{Kerber2020_ChiralMagOrder_OptExc}%
  \BibitemOpen
  \bibfield  {author} {\bibinfo {author} {\bibfnamefont {N.}~\bibnamefont
  {Kerber}}, \bibinfo {author} {\bibfnamefont {D.}~\bibnamefont {Ksenzov}},
  \bibinfo {author} {\bibfnamefont {F.}~\bibnamefont {Freimuth}}, \bibinfo
  {author} {\bibfnamefont {F.}~\bibnamefont {Capotondi}}, \bibinfo {author}
  {\bibfnamefont {E.}~\bibnamefont {Pedersoli}}, \bibinfo {author}
  {\bibfnamefont {I.}~\bibnamefont {Lopez-Quintas}}, \bibinfo {author}
  {\bibfnamefont {B.}~\bibnamefont {Seng}}, \bibinfo {author} {\bibfnamefont
  {J.}~\bibnamefont {Cramer}}, \bibinfo {author} {\bibfnamefont
  {K.}~\bibnamefont {Litzius}}, \bibinfo {author} {\bibfnamefont
  {D.}~\bibnamefont {Lacour}}, \bibinfo {author} {\bibfnamefont
  {H.}~\bibnamefont {Zabel}}, \bibinfo {author} {\bibfnamefont
  {Y.}~\bibnamefont {Mokrousov}}, \bibinfo {author} {\bibfnamefont
  {M.}~\bibnamefont {Kläui}},\ and\ \bibinfo {author} {\bibfnamefont
  {C.}~\bibnamefont {Gutt}},\ }\href
  {https://doi.org/10.1038/s41467-020-19613-z} {\bibfield  {journal} {\bibinfo
  {journal} {Nature Commun.}\ }\textbf {\bibinfo {volume} {11}},\ \bibinfo
  {pages} {6304} (\bibinfo {year} {2020})}\BibitemShut {NoStop}%
\bibitem [{\citenamefont {Ghosh}\ \emph {et~al.}(2022)\citenamefont {Ghosh},
  \citenamefont {Manchon},\ and\ \citenamefont
  {{\v{Z}}elezn{\'{y}}}}]{Ghosh2022_STT_Ncol_AFM}%
  \BibitemOpen
  \bibfield  {author} {\bibinfo {author} {\bibfnamefont {S.}~\bibnamefont
  {Ghosh}}, \bibinfo {author} {\bibfnamefont {A.}~\bibnamefont {Manchon}},\
  and\ \bibinfo {author} {\bibfnamefont {J.}~\bibnamefont
  {{\v{Z}}elezn{\'{y}}}},\ }\href
  {https://doi.org/10.1103/physrevlett.128.097702} {\bibfield  {journal}
  {\bibinfo  {journal} {Phys. Rev. Lett.}\ }\textbf {\bibinfo {volume} {128}},\
  \bibinfo {pages} {097702} (\bibinfo {year} {2022})}\BibitemShut {NoStop}%
\bibitem [{\citenamefont {Xia}\ \emph {et~al.}(2006)\citenamefont {Xia},
  \citenamefont {Zwierzycki}, \citenamefont {Talanana}, \citenamefont {Kelly},\
  and\ \citenamefont {Bauer}}]{Xia:prb06}%
  \BibitemOpen
  \bibfield  {author} {\bibinfo {author} {\bibfnamefont {K.}~\bibnamefont
  {Xia}}, \bibinfo {author} {\bibfnamefont {M.}~\bibnamefont {Zwierzycki}},
  \bibinfo {author} {\bibfnamefont {M.}~\bibnamefont {Talanana}}, \bibinfo
  {author} {\bibfnamefont {P.~J.}\ \bibnamefont {Kelly}},\ and\ \bibinfo
  {author} {\bibfnamefont {G.~E.~W.}\ \bibnamefont {Bauer}},\ }\href
  {https://doi.org/10.1103/PhysRevB.73.064420} {\bibfield  {journal} {\bibinfo
  {journal} {Phys. Rev. B}\ }\textbf {\bibinfo {volume} {73}},\ \bibinfo
  {pages} {064420} (\bibinfo {year} {2006})}\BibitemShut {NoStop}%
\bibitem [{\citenamefont {Zwierzycki}\ \emph {et~al.}(2008)\citenamefont
  {Zwierzycki}, \citenamefont {Khomyakov}, \citenamefont {Starikov},
  \citenamefont {Xia}, \citenamefont {Talanana}, \citenamefont {Xu},
  \citenamefont {Karpan}, \citenamefont {Marushchenko}, \citenamefont {Turek},
  \citenamefont {Bauer}, \citenamefont {Brocks},\ and\ \citenamefont
  {Kelly}}]{Zwierzycki2008:PSSb}%
  \BibitemOpen
  \bibfield  {author} {\bibinfo {author} {\bibfnamefont {M.}~\bibnamefont
  {Zwierzycki}}, \bibinfo {author} {\bibfnamefont {P.~A.}\ \bibnamefont
  {Khomyakov}}, \bibinfo {author} {\bibfnamefont {A.~A.}\ \bibnamefont
  {Starikov}}, \bibinfo {author} {\bibfnamefont {K.}~\bibnamefont {Xia}},
  \bibinfo {author} {\bibfnamefont {M.}~\bibnamefont {Talanana}}, \bibinfo
  {author} {\bibfnamefont {P.~X.}\ \bibnamefont {Xu}}, \bibinfo {author}
  {\bibfnamefont {V.~M.}\ \bibnamefont {Karpan}}, \bibinfo {author}
  {\bibfnamefont {I.}~\bibnamefont {Marushchenko}}, \bibinfo {author}
  {\bibfnamefont {I.}~\bibnamefont {Turek}}, \bibinfo {author} {\bibfnamefont
  {G.~E.~W.}\ \bibnamefont {Bauer}}, \bibinfo {author} {\bibfnamefont
  {G.}~\bibnamefont {Brocks}},\ and\ \bibinfo {author} {\bibfnamefont {P.~J.}\
  \bibnamefont {Kelly}},\ }\href
  {https://doi.org/https://doi.org/10.1002/pssb.200743359} {\bibfield
  {journal} {\bibinfo  {journal} {physica status solidi (b)}\ }\textbf
  {\bibinfo {volume} {245}},\ \bibinfo {pages} {623} (\bibinfo {year}
  {2008})}\BibitemShut {NoStop}%
\bibitem [{\citenamefont {Bass}\ and\ \citenamefont {{Pratt
  Jr.}}(1999)}]{Bass:jmmm99}%
  \BibitemOpen
  \bibfield  {author} {\bibinfo {author} {\bibfnamefont {J.}~\bibnamefont
  {Bass}}\ and\ \bibinfo {author} {\bibfnamefont {W.~P.}\ \bibnamefont {{Pratt
  Jr.}}},\ }\href {https://doi.org/10.1016/S0304-8853(99)00316-9} {\bibfield
  {journal} {\bibinfo  {journal} {J. Magn. \& Magn. Mater.}\ }\textbf {\bibinfo
  {volume} {200}},\ \bibinfo {pages} {274} (\bibinfo {year}
  {1999})}\BibitemShut {NoStop}%
\bibitem [{\citenamefont {Bass}\ and\ \citenamefont {{Pratt
  Jr.}}(2007)}]{Bass:jpcm07}%
  \BibitemOpen
  \bibfield  {author} {\bibinfo {author} {\bibfnamefont {J.}~\bibnamefont
  {Bass}}\ and\ \bibinfo {author} {\bibfnamefont {W.~P.}\ \bibnamefont {{Pratt
  Jr.}}},\ }\href {https://doi.org/10.1088/0953-8984/19/18/183201} {\bibfield
  {journal} {\bibinfo  {journal} {J. Phys.: Condens. Matter}\ }\textbf
  {\bibinfo {volume} {19}},\ \bibinfo {pages} {183201} (\bibinfo {year}
  {2007})}\BibitemShut {NoStop}%
\bibitem [{\citenamefont {Lu}\ \emph {et~al.}(2020)\citenamefont {Lu},
  \citenamefont {Zhao}, \citenamefont {Battiato}, \citenamefont {Wu},\ and\
  \citenamefont {Yuan}}]{Lu:PRB2020}%
  \BibitemOpen
  \bibfield  {author} {\bibinfo {author} {\bibfnamefont {W.-T.}\ \bibnamefont
  {Lu}}, \bibinfo {author} {\bibfnamefont {Y.}~\bibnamefont {Zhao}}, \bibinfo
  {author} {\bibfnamefont {M.}~\bibnamefont {Battiato}}, \bibinfo {author}
  {\bibfnamefont {Y.}~\bibnamefont {Wu}},\ and\ \bibinfo {author}
  {\bibfnamefont {Z.}~\bibnamefont {Yuan}},\ }\href
  {https://doi.org/10.1103/PhysRevB.101.014435} {\bibfield  {journal} {\bibinfo
   {journal} {Phys. Rev. B}\ }\textbf {\bibinfo {volume} {101}},\ \bibinfo
  {pages} {014435} (\bibinfo {year} {2020})}\BibitemShut {NoStop}%
\bibitem [{\citenamefont {Battiato}\ \emph {et~al.}(2014)\citenamefont
  {Battiato}, \citenamefont {Maldonado},\ and\ \citenamefont
  {Oppeneer}}]{Battiato2014_JAP}%
  \BibitemOpen
  \bibfield  {author} {\bibinfo {author} {\bibfnamefont {M.}~\bibnamefont
  {Battiato}}, \bibinfo {author} {\bibfnamefont {P.}~\bibnamefont
  {Maldonado}},\ and\ \bibinfo {author} {\bibfnamefont {P.~M.}\ \bibnamefont
  {Oppeneer}},\ }\href {https://doi.org/10.1063/1.4870589} {\bibfield
  {journal} {\bibinfo  {journal} {J. Appl. Phys.}\ }\textbf {\bibinfo {volume}
  {115}},\ \bibinfo {pages} {172611} (\bibinfo {year} {2014})}\BibitemShut
  {NoStop}%
\bibitem [{\citenamefont {Nenno}\ \emph {et~al.}(2016)\citenamefont {Nenno},
  \citenamefont {Kaltenborn},\ and\ \citenamefont {Schneider}}]{Nenno2016}%
  \BibitemOpen
  \bibfield  {author} {\bibinfo {author} {\bibfnamefont {D.~M.}\ \bibnamefont
  {Nenno}}, \bibinfo {author} {\bibfnamefont {S.}~\bibnamefont {Kaltenborn}},\
  and\ \bibinfo {author} {\bibfnamefont {H.~C.}\ \bibnamefont {Schneider}},\
  }\href {https://doi.org/10.1103/PhysRevB.94.115102} {\bibfield  {journal}
  {\bibinfo  {journal} {Phys. Rev. B}\ }\textbf {\bibinfo {volume} {94}},\
  \bibinfo {pages} {115102} (\bibinfo {year} {2016})}\BibitemShut {NoStop}%
\bibitem [{\citenamefont {Nenno}\ \emph {et~al.}(2018)\citenamefont {Nenno},
  \citenamefont {Rethfeld},\ and\ \citenamefont {Schneider}}]{Nenno2018}%
  \BibitemOpen
  \bibfield  {author} {\bibinfo {author} {\bibfnamefont {D.~M.}\ \bibnamefont
  {Nenno}}, \bibinfo {author} {\bibfnamefont {B.}~\bibnamefont {Rethfeld}},\
  and\ \bibinfo {author} {\bibfnamefont {H.~C.}\ \bibnamefont {Schneider}},\
  }\href {https://doi.org/10.1103/PhysRevB.98.224416} {\bibfield  {journal}
  {\bibinfo  {journal} {Phys. Rev. B}\ }\textbf {\bibinfo {volume} {98}},\
  \bibinfo {pages} {224416} (\bibinfo {year} {2018})}\BibitemShut {NoStop}%
\bibitem [{\citenamefont {Zwierzycki}\ \emph {et~al.}(2005)\citenamefont
  {Zwierzycki}, \citenamefont {Tserkovnyak}, \citenamefont {Kelly},
  \citenamefont {Brataas},\ and\ \citenamefont {Bauer}}]{Zwierzycki2005:PRB}%
  \BibitemOpen
  \bibfield  {author} {\bibinfo {author} {\bibfnamefont {M.}~\bibnamefont
  {Zwierzycki}}, \bibinfo {author} {\bibfnamefont {Y.}~\bibnamefont
  {Tserkovnyak}}, \bibinfo {author} {\bibfnamefont {P.~J.}\ \bibnamefont
  {Kelly}}, \bibinfo {author} {\bibfnamefont {A.}~\bibnamefont {Brataas}},\
  and\ \bibinfo {author} {\bibfnamefont {G.~E.~W.}\ \bibnamefont {Bauer}},\
  }\href {https://doi.org/10.1103/PhysRevB.71.064420} {\bibfield  {journal}
  {\bibinfo  {journal} {Phys. Rev. B}\ }\textbf {\bibinfo {volume} {71}},\
  \bibinfo {pages} {064420} (\bibinfo {year} {2005})}\BibitemShut {NoStop}%
\bibitem [{\citenamefont {Andersen}\ \emph {et~al.}(1985)\citenamefont
  {Andersen}, \citenamefont {Jepsen},\ and\ \citenamefont
  {Gl{\"{o}}tzel}}]{Andersen:85}%
  \BibitemOpen
  \bibfield  {author} {\bibinfo {author} {\bibfnamefont {O.~K.}\ \bibnamefont
  {Andersen}}, \bibinfo {author} {\bibfnamefont {O.}~\bibnamefont {Jepsen}},\
  and\ \bibinfo {author} {\bibfnamefont {D.}~\bibnamefont {Gl{\"{o}}tzel}},\
  }in\ \href@noop {} {\emph {\bibinfo {booktitle} {Highlights of Condensed
  Matter Theory}}},\ \bibinfo {series and number} {International School of
  Physics `Enrico Fermi', Varenna, Italy,},\ \bibinfo {editor} {edited by\
  \bibinfo {editor} {\bibfnamefont {F.}~\bibnamefont {Bassani}}, \bibinfo
  {editor} {\bibfnamefont {F.}~\bibnamefont {Fumi}},\ and\ \bibinfo {editor}
  {\bibfnamefont {M.~P.}\ \bibnamefont {Tosi}}}\ (\bibinfo  {publisher}
  {North-Holland},\ \bibinfo {address} {Amsterdam},\ \bibinfo {year} {1985})\
  pp.\ \bibinfo {pages} {59--176}\BibitemShut {NoStop}%
\bibitem [{\citenamefont {Turek}\ \emph {et~al.}(1997)\citenamefont {Turek},
  \citenamefont {Drchal}, \citenamefont {Kudrnovsk\'{y}}, \citenamefont
  {\v{S}ob},\ and\ \citenamefont {Weinberger}}]{Turek:97}%
  \BibitemOpen
  \bibfield  {author} {\bibinfo {author} {\bibfnamefont {I.}~\bibnamefont
  {Turek}}, \bibinfo {author} {\bibfnamefont {V.}~\bibnamefont {Drchal}},
  \bibinfo {author} {\bibfnamefont {J.}~\bibnamefont {Kudrnovsk\'{y}}},
  \bibinfo {author} {\bibfnamefont {M.}~\bibnamefont {\v{S}ob}},\ and\ \bibinfo
  {author} {\bibfnamefont {P.}~\bibnamefont {Weinberger}},\ }\href@noop {}
  {\emph {\bibinfo {title} {Electronic Structure of Disordered Alloys, Surfaces
  and Interfaces}}}\ (\bibinfo  {publisher} {Kluwer},\ \bibinfo {address}
  {Boston-London-Dordrecht},\ \bibinfo {year} {1997})\BibitemShut {NoStop}%
\bibitem [{\citenamefont {Bal\'a\ifmmode~\check{z}\else \v{z}\fi{}}\ \emph
  {et~al.}(2013)\citenamefont {Bal\'a\ifmmode~\check{z}\else \v{z}\fi{}},
  \citenamefont {Zwierzycki},\ and\ \citenamefont {Barna\ifmmode~\acute{s}\else
  \'{s}\fi{}}}]{Balaz_prb13}%
  \BibitemOpen
  \bibfield  {author} {\bibinfo {author} {\bibfnamefont {P.}~\bibnamefont
  {Bal\'a\ifmmode~\check{z}\else \v{z}\fi{}}}, \bibinfo {author} {\bibfnamefont
  {M.}~\bibnamefont {Zwierzycki}},\ and\ \bibinfo {author} {\bibfnamefont
  {J.}~\bibnamefont {Barna\ifmmode~\acute{s}\else \'{s}\fi{}}},\ }\href
  {https://doi.org/10.1103/PhysRevB.88.094422} {\bibfield  {journal} {\bibinfo
  {journal} {Phys. Rev. B}\ }\textbf {\bibinfo {volume} {88}},\ \bibinfo
  {pages} {094422} (\bibinfo {year} {2013})}\BibitemShut {NoStop}%
\bibitem [{\citenamefont {Yastremsky}\ \emph {et~al.}(2014)\citenamefont
  {Yastremsky}, \citenamefont {Oppeneer},\ and\ \citenamefont
  {Ivanov}}]{Yastremsky2014}%
  \BibitemOpen
  \bibfield  {author} {\bibinfo {author} {\bibfnamefont {I.~A.}\ \bibnamefont
  {Yastremsky}}, \bibinfo {author} {\bibfnamefont {P.~M.}\ \bibnamefont
  {Oppeneer}},\ and\ \bibinfo {author} {\bibfnamefont {B.~A.}\ \bibnamefont
  {Ivanov}},\ }\href {https://doi.org/10.1103/PhysRevB.90.024409} {\bibfield
  {journal} {\bibinfo  {journal} {Phys. Rev. B}\ }\textbf {\bibinfo {volume}
  {90}},\ \bibinfo {pages} {024409} (\bibinfo {year} {2014})}\BibitemShut
  {NoStop}%
\bibitem [{\citenamefont {Stiles}\ and\ \citenamefont
  {Zangwill}(2002)}]{r_02_StilesAnatomy}%
  \BibitemOpen
  \bibfield  {author} {\bibinfo {author} {\bibfnamefont {M.~D.}\ \bibnamefont
  {Stiles}}\ and\ \bibinfo {author} {\bibfnamefont {A.}~\bibnamefont
  {Zangwill}},\ }\href {https://link.aps.org/doi/10.1103/PhysRevB.66.014407}
  {\bibfield  {journal} {\bibinfo  {journal} {Phys. Rev. B}\ }\textbf {\bibinfo
  {volume} {66}},\ \bibinfo {pages} {014407} (\bibinfo {year}
  {2002})}\BibitemShut {NoStop}%
\bibitem [{\citenamefont {Ghosh}\ \emph {et~al.}(2012)\citenamefont {Ghosh},
  \citenamefont {Auffret}, \citenamefont {Ebels},\ and\ \citenamefont
  {Bailey}}]{Ghosh2012:PRL}%
  \BibitemOpen
  \bibfield  {author} {\bibinfo {author} {\bibfnamefont {A.}~\bibnamefont
  {Ghosh}}, \bibinfo {author} {\bibfnamefont {S.}~\bibnamefont {Auffret}},
  \bibinfo {author} {\bibfnamefont {U.}~\bibnamefont {Ebels}},\ and\ \bibinfo
  {author} {\bibfnamefont {W.~E.}\ \bibnamefont {Bailey}},\ }\href
  {https://doi.org/10.1103/PhysRevLett.109.127202} {\bibfield  {journal}
  {\bibinfo  {journal} {Phys. Rev. Lett.}\ }\textbf {\bibinfo {volume} {109}},\
  \bibinfo {pages} {127202} (\bibinfo {year} {2012})}\BibitemShut {NoStop}%
\bibitem [{\citenamefont {Barna{\'{s}}}\ \emph {et~al.}(2005)\citenamefont
  {Barna{\'{s}}}, \citenamefont {Fert}, \citenamefont {Gmitra}, \citenamefont
  {Weymann},\ and\ \citenamefont {Dugaev}}]{Barnas2005}%
  \BibitemOpen
  \bibfield  {author} {\bibinfo {author} {\bibfnamefont {J.}~\bibnamefont
  {Barna{\'{s}}}}, \bibinfo {author} {\bibfnamefont {A.}~\bibnamefont {Fert}},
  \bibinfo {author} {\bibfnamefont {M.}~\bibnamefont {Gmitra}}, \bibinfo
  {author} {\bibfnamefont {I.}~\bibnamefont {Weymann}},\ and\ \bibinfo {author}
  {\bibfnamefont {V.~K.}\ \bibnamefont {Dugaev}},\ }\href
  {https://link.aps.org/doi/10.1103/PhysRevB.72.024426} {\bibfield  {journal}
  {\bibinfo  {journal} {Phys. Rev. B}\ }\textbf {\bibinfo {volume} {72}},\
  \bibinfo {pages} {024426} (\bibinfo {year} {2005})}\BibitemShut {NoStop}%
\end{thebibliography}%


%

\end{document}